\documentclass{article}
\usepackage{amsfonts}
\usepackage{amsmath}
\usepackage{amssymb}
\usepackage{graphicx}
\usepackage{float}

\def\be{\begin{eqnarray}}
\def\ee{\end{eqnarray}}

\def\bfig{\begin{figure}[H] }
\def\efig{\end{figure}}

\def\bc{\begin{center}}
\def\ec{\end{center}}

\def\nn{\nonumber}
\def\p{\partial}
\def\tr{{\rm tr}}

\hoffset= - 1.0in         % switch off for draft style
\begin{document}

\hfill ITEP/TH-02/10

\bigskip

\centerline{\Large{Chern-Simons Theory in the Temporal Gauge and
}} \centerline{\Large{Knot Invariants through the Universal
Quantum R-Matrix }}

\bigskip

\centerline{\it Alexei Morozov and Andrey Smirnov}

\bigskip

\centerline{ITEP, Moscow, Russia}

\bigskip

\centerline{ABSTRACT}

\bigskip

{\footnotesize In temporal gauge $A_0=0$ the $3d$ Chern-Simons
theory acquires quadratic action and an ultralocal propagator.
This directly implies a $2d$ $R$-matrix representation for the
correlators of Wilson lines (knot invariants), where only the
crossing points of the contours projection on the $xy$ plane
contribute. Though the theory is quadratic, $P$-exponents remain
non-trivial operators and $R$-factors are easier to guess then
derive. We show that the topological invariants arise if
additional flag structure $\mathbb{R}^3 \supset \mathbb{R}^2
\supset \mathbb{R}^1$ ($xy$ plane and an $y$ line in it) is
introduced, $R$ is the universal quantum $R$-matrix and turning
points contribute the "enhancement" factors $q^{\,\rho}$. }

\bigskip

\tableofcontents

\section{Introduction \label{intro}}

Since the seminal work of E.Witten \cite{WCS} (see also
\cite{Atiyah}) it is known that knot invariants should be
described as correlators of Wilson lines in $3d$ Chern-Simons
theory (CST), provided they are somehow regularized in a way,
which preserves topological invariance of the classical CST
\cite{AlS}. Enormous efforts were applied since then
\cite{Bar1}-\cite{Ka4} in order to realize this idea and associate
various existing representations of knot invariants
\cite{Reidem}-\cite{katlas} with different gauge choices in CST.
Detailed overview of entire situation and references is beyond the
scope of this paper, where we are going to concentrate on one
distinguished implication of the Witten's conjecture.
%Obviously \cite{Gu}-\cite{La},
Namely, it is clear that CS theory with the action \be S_{CS} = k
\int \tr \left(AdA + \frac{2}{3}A^3\right) d^3x = k
\epsilon^{\mu\nu\lambda} \int \left(A_\mu^a \p_\nu A^a_\lambda +
\frac{2}{3}f_{abc} A^a_\mu A^b_\nu A^c_\lambda\right)dt dx dy
\label{CSA} \ee becomes exactly solvable not in a transcendental,
but in straightforward and constructive way in the temporal gauge
$A_0^a=0$, when the cubic term disappears and the action becomes
quadratic \cite{La5, MR, Sm}. This free-field representation of
CST is the most natural from the point of view of QFT and string
theory \cite{str}, and it immediately implies the $R$-matrix kind
of representation of knot invariants {\it a la} \cite{Rm1}.

Amusingly, even this description remains unexplored in exhaustive
way, despite a number of promising attempts
\cite{Gu1}-\cite{La10}, see \cite{Sm} for a recent review. The
problem is not so simple because the relevant observables --
Wilson $P$-exponents -- are in no way simplified in the temporal
gauge and various regularization problems still need to be
resolved. The best way to proceed here, as usual, is to first {\it
guess} the answer, which possesses all the features it should have
in the temporal gauge, and then explain how this answer can be
technically deduced from original theory (by developing a $3d$
analogue of the $2d$ free-field calculus \cite{GMMOS} for the WZNW
model).

The present paper is devoted to the first step -- the "educated
guess". The "obvious" properties the answer should have are:

-- it is formulated in terms of projection of original $3d$
contour on the $xy$ plane,

-- it gets contributions only from the crossing and turning points
of this projection,

-- each contribution should arise in a {\it universal} --
representation-independent -- form, i.e. contain only the
generators $T_R^a$ of the algebra $G$ (or its deformation) in
representation $R$,

-- this contribution should be some clever regularization
(deformation) of the naive expression $\prod_a q^{T^a_R\otimes
T^a_R}$ with  $q = \exp\frac{2\pi i}{k_{1}}$ where $k_{1}$ is the
renormalization if the bare constant $k$.

-- the full answer should be topological invariant, i.e. invariant
under the Reidemeister moves of the projected contours, and
independent of the choice of auxiliary direction on the $xy$
plane, needed to define the "turning points".

Actually an answer with these properties exists and its
ingredients are well known, see, for example,
\cite{MV1}-\cite{MV3}: it involves the quantum universal
$R$-matrix \cite{Bax}-\cite{Drin2}, associated with the quantum
version $G_q$ of the original group $G$, and the relevant
generators $T^a_R$ are those of the quantum algebra
\cite{qalg1}-\cite{qalg4}. The only problem is that this
construction is naturally associated with the operators, which are
group elements of $G_q$ \cite{MV2}, rather than the Wilson
$P$-exponents in the temporal gauge. Thus, if one believes in this
answer, the next step should be the explanation how the group
elements arise in "bosonization" of Wilson loops. This step,
however, will be left for the future research, while in the
present text we just add a little more to the motivation in
s.\ref{motiv}, find equations that describes topological
properties of vevs of Wilson loops in the temporal gauge in
s.\ref{top} and finally, describe their possible solution in
s.\ref{Sol}. We also give several examples of explicit
computations of knot invariants in s.\ref{Ex}.  This paper can be
considered as completion of the guessing process, originated long
ago in \cite{Gu8,Gu9} and \cite{MR}.

%\vspace{7mm}
\section{Motivation \label{motiv}}

In the $A_0=0$ gauge the CS action (\ref{CSA}) becomes quadratic,
\be S_{CS} = k \int (A_x^a\dot A_y^a - A_y^a\dot A_x^a) dt dx dy
\ee and the propagator becomes \be \left< A_i^a(\vec x),
A_j^b(\vec x')\right> = \delta^{ab} \epsilon_{ij} {\rm sign}(t-t')
\delta^{(2)}(\vec x-\vec x') \label{prop} \ee This means that the
correlator of the Wilson lines in representation $\varrho$ of the
gauge group $G$ \be W_\varrho(C) = \tr_\varrho P\exp \oint_{C}
A_\mu dx^\mu = \tr_\varrho P\exp \oint_{\tilde C} (A_xdx + A_ydy)
\label{Wili} \ee gets contribution only from the
(self)-intersections of the contours $\tilde C_i$, which are the
$2d$ projections of the $3d$ contours $C_i$ onto the $xy$ plane.

Let us pick up a direction in the $xy$ plane, and let it be the
$y$ axis. Then instead of the closed Wilson lines we can consider
the matrix (representation)-valued "open" lines \be U(\tilde C, y)
= P\exp \int^y_{\tilde C} (A_x dx + A_y dy) \label{Uli} \ee and
study the "evolution" of the ordered product
$\prod\limits^\rightarrow U(\tilde C_i,y)$ with the change of $y$.
Actually there is no evolution, the product does not depend on
$y$, except for the "moments" when some two lines intersect
(crossing points) or some line turns backwards (turning point).
Then some factors $\cal R$ and $\cal Q$ appear, acting in the
product of two representations in the first case and in the single
representation in the second case. The factor ${\cal R}$ is
independent of the angle between the intersecting lines, this
property can be observed  in the Abelian situation, see \cite{Sm}
for definiteness. Denote the state at the slice (moment) $1$ in
Fig.\ref{evo}(a) by $U\otimes V$. \bfig \bc
\includegraphics[scale=0.75]{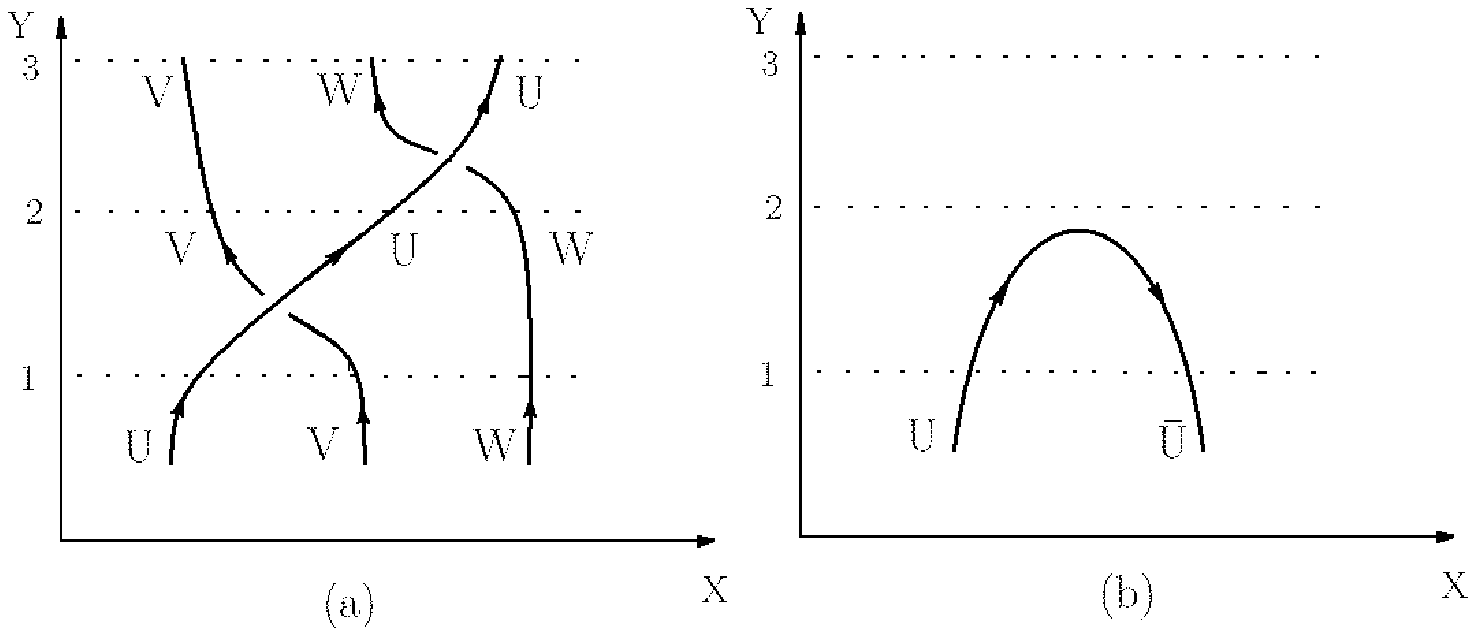} \caption{\label{evo}} \ec
\efig \noindent Then at the moment $2$ it will turn into ${\cal
R}_{UV} (U\otimes V)$. At the same time it should be something
made from reordered product $V\otimes U$. The natural guess is \be
{\cal R}_{UV} (U\otimes V) = (V\otimes U){\cal R}_{UV} \label{RUV}
\ee and, looking at the propagator (\ref{prop}), we understand
that \be {\cal R}_{UV} \approx \left(q^{T_R^a\otimes
T_R^a}\right)_{\rm reg} \label{qTT} \ee with $q=\exp(2\pi
i/k_{1})$ and somehow deformed r.h.s., because regularization can
and do modify the naive answer. In the last expression  Similarly,
if the third operator $\bar W$ is involved, then we get \be
U\otimes V \otimes \bar W & {\rm at\ the\ slice}\ 1, \nn \\
{\cal R}_{UV}(U\otimes V \otimes \bar W) = (V\otimes U \otimes\bar
W) {\cal R}_{UV}
& {\rm at\ the\ slice}\ 2, \\
{\cal R}_{U\bar W}{\cal R}_{UV} (U\otimes V \otimes \bar W) =
{\cal R}_{U\bar W}(V\otimes U \otimes\bar W) {\cal R}_{UV} =
(V\otimes \bar W \otimes U) {\cal R}_{U\bar W}{\cal R}_{UV} & {\rm
at\ the\ slice}\ 3, \nn \ee and so on. At the turning point we
have a transition Fig.\ref{evo}(b) \be {\cal Q}_U (U\otimes \bar
U) = 1 \ee from a pair of operators into an empty state.

Entire knot can be described as a transition from an empty state
at $y = -\infty$ to an empty state at $y=\infty$ with the
associated amplitude, made from a product of ${\cal R}$ and ${\cal
Q}$ factors. For this product to transform under Reidemeister
moves as vev of a Wilson loop the elementary factors should
satisfy some bilinear and trilinear relations see s.\ref{top}, the
main one being the Yang-Baxter equation for ${\cal R}$. Then
(\ref{qTT}) implies that ${\cal R}$ is the quantum universal
$R$-matrix of $G_q$ \cite{MV1}-\cite{qalg4}. For such choice of
${\cal R}$ the operators $U,V,W$ with the property (\ref{RUV}) do
indeed exist: the simplest example being the group elements of
$G_q$ \cite{MV2}. As explained in the introduction, it remains to
understand why the regularized open Wilson lines (\ref{Uli}) have
the same commutation relations (\ref{qTT}) -- and this is beyond
the scope of the present paper. Note that understanding of this
fact is important for Chern-Simons theory, but is not really
needed for building knot invariants: just instead of referring to
the "obvious" topological invariance of Wilson-line correlators in
CST one needs to prove invariance of the product explicitly - and
this will be done explicitly in s.\ref{Sol} below.

\section{Operators arising from CST in temporal gauge \label{top} }
The invariance of vevs of Wilson loops under topological
deformations imposes several conditions on the operators arising
from CST in the temporal gauge. The purpose of this section is to
argue that there are only two independent operators from which the
vevs of Wilson loops can be constructed. We also derive the
equations for these operators that ensures the topological
invariance of vevs.
\subsection{Projected knots}
If additional flag structure $ \mathbb{R}^{3} \supset \mathbb{R}^2
\supset \mathbb{R}^1$ on the space is fixed, then according to
sections \ref{intro} and \ref{motiv} we suppose that the answer
for vev of a Wilson loop has the following properties:
\begin{itemize}
\item The answer is formulated in terms of a projection of a knot
to a fixed plane $\mathbb{R}^2$. For definiteness,  we specify
this plane to be $x y$-plane in some reference frame. \item The
answer gets angle-independent contributions only from crossings of
the resulting two-dimensional curve and its turning points with
respect to the fixed direction $\mathbb{R}$ in the plane. We
specify this direction to be $y$-direction in our reference frame.
This means that we need the universally defined operators
corresponding to the crossings and the turning points. The vevs of
Wilson loops are just appropriate contractions of these operators.
 \item Under topological
deformations these contractions are transformed exactly as vevs of
Wilson loops in~CST.
\end{itemize}
With these desired properties in mind, let us consider a
two-dimensional projection of a knot to the fixed $x y$ plane, for
example as in the Fig.\ref{XY}($a$) .  The arrows denote the
orientation of the knot. Without loss of generality, we assume
that all crossings of the resulting two-dimensional curve have two
upper and two lower legs about fixed $y$-direction.  For example
in the Fig.\ref{XY}($b$) $a$, $b$ are upper and $c$, $d$ are lower
legs. Obviously, every crossing can be reduced to this form by a
small deformation in the plane. \bfig \bc
\includegraphics[scale=0.7]{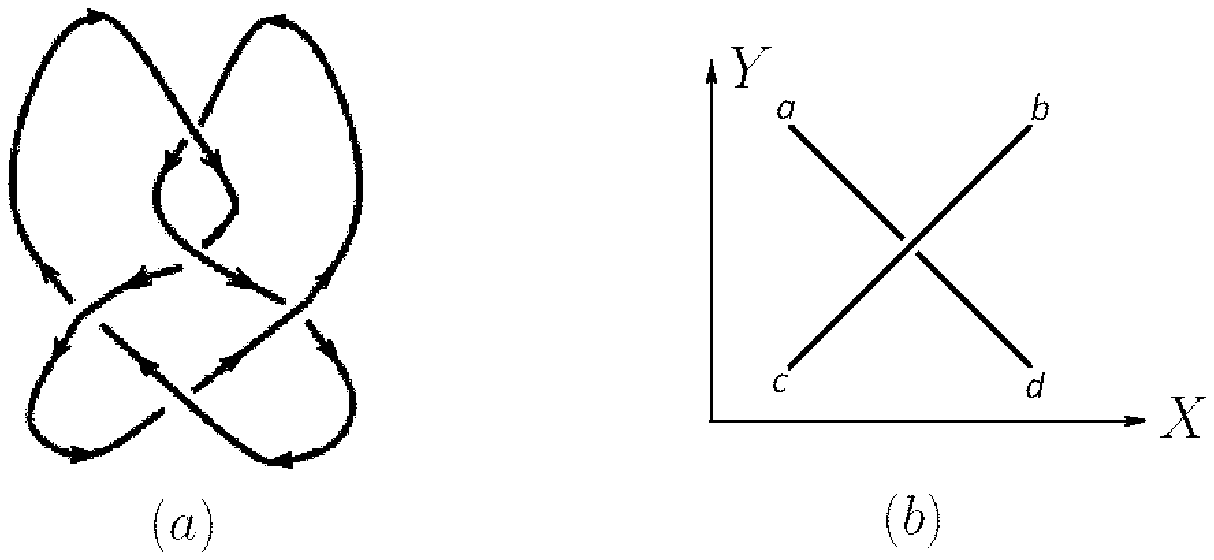}  \caption{\label{XY}} \ec
\efig
 We can see that for every two-dimensional
projection only $8$ types of crossings, which differ from each
other by orientation and $4$ types of turning points about $y$
direction can appear (see Fig.\ref{CT}). \bfig \bc
\includegraphics[scale=0.7]{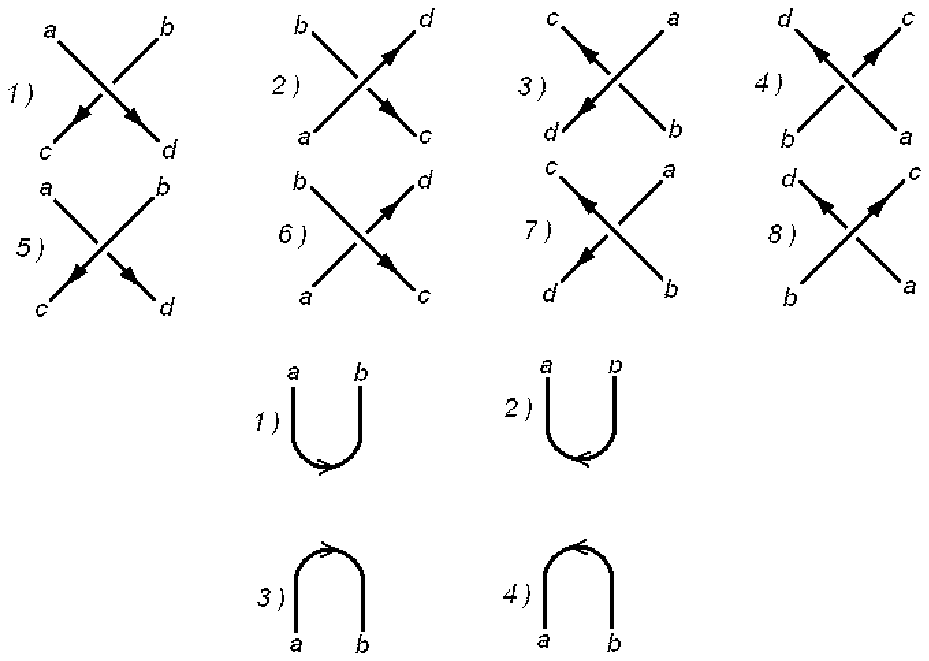}  \caption{\label{CT}} \ec
\efig
 According to the second property from of our list, we
define $8$ corresponding crossing operators ${{\cal
R}_{i}}^{\,a\,b}_{\,c\,d}, \ \ i=1...8$ and $4$ operators for each
type of turning point ${{\cal M}_{j}}^{\,a}_{\,b}, \ \ j=1...4$.
The indices of these operators correspond to free legs of the
crossings and the turning points in the two-dimensional picture.
We assume that the upper index denotes an incoming leg and the
lower denotes an outgoing one. Therefore, the crossing operators
have two upper and two lower indices, the turning point operators
have one index of each type Fig.\ref{Rm8}. \bfig \bc
\includegraphics[scale=0.8]{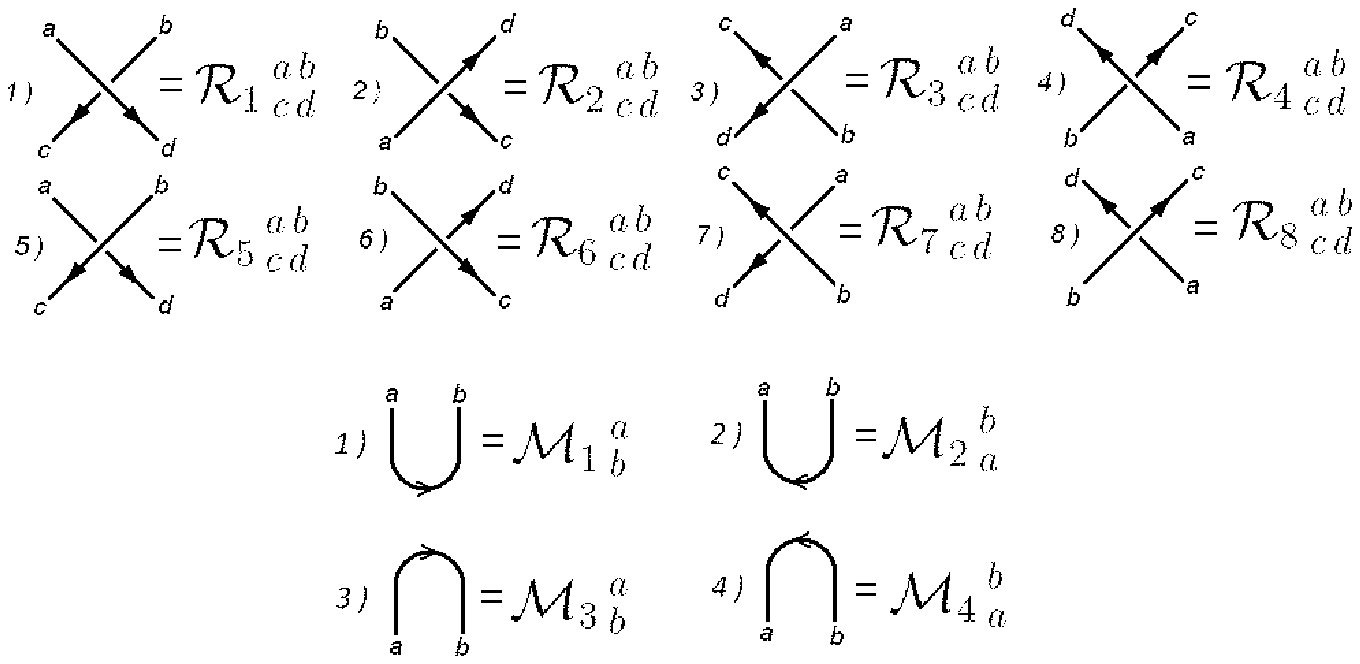}  \caption{\label{Rm8}} \ec
\efig
  The vev of a Wilson loop $\,<W(c)>\,$ can be represented
in the form of contractions of these operators. More precisely, to
find the vev $<W(c)>$, we need to peak a two dimensional
projection of the closed curve $c$, then attach to every crossing
and turning points corresponding operators and contract the
indices according to the two dimensional diagram. For example, for
unknot $U_{0}$ with the two-dimensional projection represented in
the Fig.\ref{Un} the result is: \be <W(U_{0})>\,={{\cal
M}_3}^{a}_{b}\,{{\cal M}_1}^{b}_{c}\,{{\cal M}_3}^{c}_{d}\,{{\cal
M}_2}^{d}_{a}\,=\tr( {{\cal M}_3}{{\cal M}_1} {{\cal M}_3}{{\cal
M}_2}) \ee
 \bfig \bc
\includegraphics[scale=0.8]{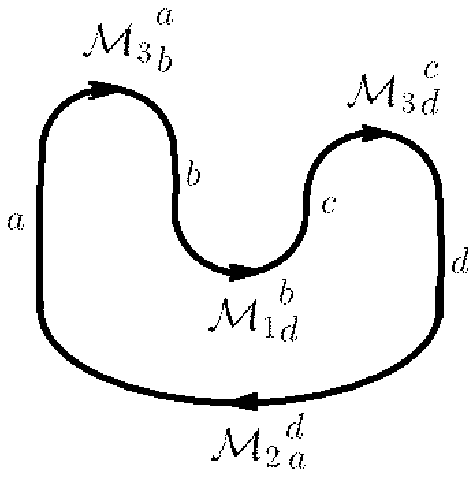}  \caption{\label{Un}} \ec
\efig
  The CST implies that the
result does not depend on the choice of the projection and is
invariant under smooth topological deformations. This property
imposes several conditions on the crossing and turning point
operators. In the rest of this section we analyze these
restrictions and derive the fundamental equations (\ref{fundeq})
for the special point operators.  These equations ensures that
under topological deformations the resulting contractions of the
operators have the same transformation properties as the vevs of
Wilson loops. %\vspace{7mm}
\subsection{Turning point operators}
Simple topological consideration shows, that  the four operators
${{\cal M}}_{j},\ \ j=1..4$, are not independent. Indeed, the
straight vertical line is topologically equivalent to a curved
line with two turning points Fig.\ref{turn}.  Then the topological
invariance of vevs of Wilson loops in CST implies that the
corresponding turning point operators are related by the
inversion: \be \label{turrel }{\cal M}_{3}={\cal M}_{1}^{-1}\ \ \
{\cal M}_{2}={\cal M}_{4}^{-1} \ee \bfig \bc
\includegraphics[scale=0.8]{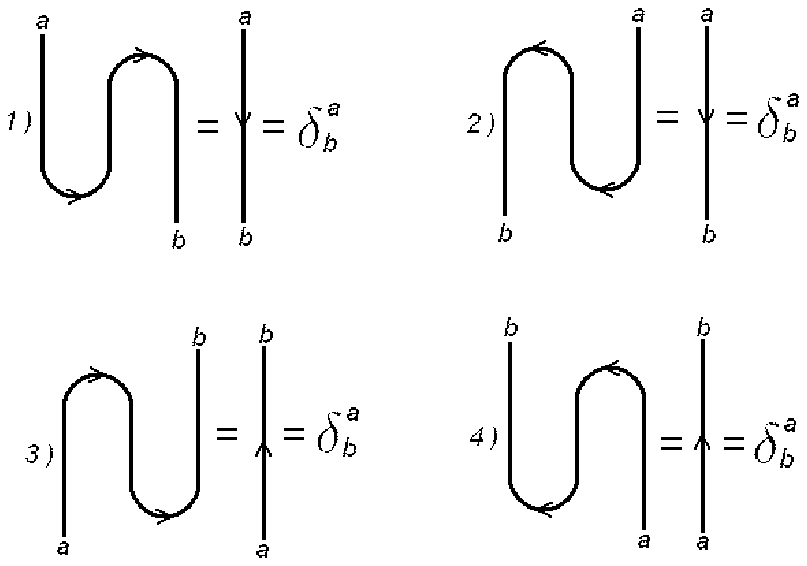}  \caption{\label{turn}} \ec
\efig \noindent Therefore, in fact, there are only two independent
turning point operators ${\cal M}$ and  $\overline{{\cal M}}$, and
the others can be expressed trough them: \be \label{tur} {\cal
M}_{1}={\cal M}, \ \ {\cal M}_{4}=\overline{{\cal M}},\ \ {\cal
M}_{3}={\cal M}^{-1},\ \ {\cal M}_{2}=\overline{{\cal M}}^{\,-1}
\ee %\vspace{7mm}
\subsection{Crossing operators}
Similar topological arguments show that there are only two
independent crossing operators. Indeed, let us consider the
topological deformations represented in the Fig.\ref{relat}. The
topological invariance of vevs of Wilson loops in CST implies that
the crossing operators ${{\cal R}}_{2}, {{\cal R}}_{3}$ and
${{\cal R}}_{4}$ can be expressed trough the first one ${{\cal
R}}_{1}:={\cal R}$: \be  \label{crosspl}\nn {{\cal
R}_{2}}^{a\,b}_{c\,d}={{\cal
R}}^{m\,b}_{c\,k}\,{{\cal M}^{-1}}^{a}_{m}\,{{\cal M}}^{k}_{d}\ \ \ \ \Rightarrow \ \ \ \ \ {\cal R}_{2}= (1\otimes{\cal M})\,{\cal R}\,({\cal M}\otimes 1)^{-1}  \\
{{\cal R}_{3}}^{a\,b}_{c\,d}={{\cal R}}^{a\,k}_{m\,d}\,{{\cal
\overline{M}}^{-1}}^{b}_{k}\,{\overline{{\cal M}}}^{m}_{c}  \ \ \
\ \  \Rightarrow\ \ \ \ \
{\cal R}_{3}= (\overline{{\cal M}} \otimes 1)\,{\cal R}\,(1 \otimes \overline{{\cal M}})^{-1}  \\
\nn {{\cal R}_{4}}^{a\,b}_{c\,d}={{\cal R}}^{h\,f}_{k\,i}\,{{\cal
M}^{-1}}^{a}_{h}\,{{\cal M}^{-1}}^{b}_{f}\,{{\cal
M}}^{i}_{d}\,{{\cal M}}^{k}_{c}  \ \  \Rightarrow \ \ \ \ {\cal
R}_{4}= ({\cal M} \otimes{\cal M})\,{\cal R}\,({\cal
M}\otimes{\cal M})^{-1} \ee \bfig \bc
\includegraphics[scale=0.8]{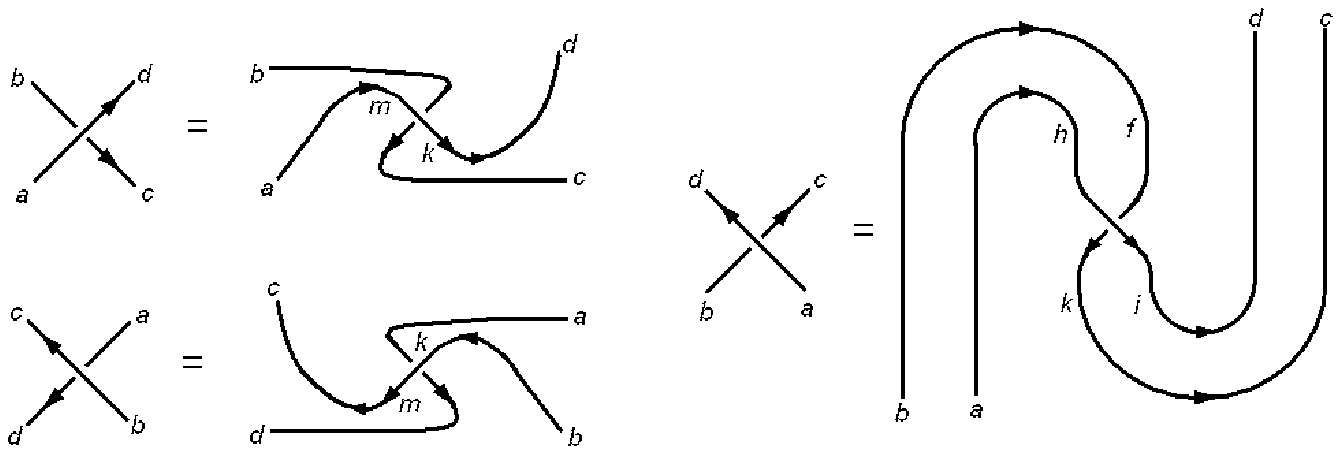}  \caption{\label{relat}} \ec
\efig \noindent If in Fig.\ref{relat} under-crossings are changed
to over-crossings then we have:
\be \label{crossmin} \nn  {\cal R}_{6}= (1\otimes{\cal M})\,\overline{{\cal R}}\,({\cal M}\otimes 1)^{-1}  \\
{\cal R}_{7}= (\overline{{\cal M}} \otimes 1)\,\overline{{\cal R}}\,(1 \otimes \overline{{\cal M}})^{-1}  \\
\nn {\cal R}_{8}= ({\cal M} \otimes{\cal M})\,\overline{{\cal
R}}\,({\cal M}\otimes{\cal M})^{-1} \ee  i.e. the crossing
operators ${\cal R}_{6}, {\cal R}_{7}$ and ${\cal R}_{8}$ can be
expressed trough ${\cal R}_{5}:=\overline{{\cal R}}$. \vspace{7mm}
\subsection{Reidemeister movies. Ambient topological
invariants.} Two knots in $\mathbb{R}^3$ are considered to be
equivalent if one of them can be transformed into the other via
smooth deformations in $\mathbb{R}^{3}$. This is the definition of
\textit{ambient isotopy equivalence} and two such knots are
referred to as \textit{ambient isotopy equivalent}. Relations
(\ref{tur})-(\ref{crossmin}) for crossing and turning point
operators ensure that the vevs constructed from them do not depend
on homotopic deformations of two-dimensional projection in the
plane, i.e. on smooth deformations in the plane that do not change
the total number of crossings of this projection. This is not
enough, however, for the vev to be topological invariant. Indeed,
different two-dimensional projections of some knot can have
different topologies, for example different numbers of crossings.
Therefore, different, non-homotopic two-dimensional projections
can represent the same knot.

Fortunately, there is a simple way to find out if the two knots
represented by their two-dimensional projections are ambient
isotopy equivalent or not \cite{Reidem}: \vspace{7mm}
\\
 \textbf{Theorem (Reidemeister)}\\ \textit{Two knots in
$\mathbb{R}^{3}$ represented by their two-dimensional projections
$P_{1}$ and $P_{2}$ are ambient isotopy equivalent if and only if
$P_{1}$ can be deformed into $P_{2}$ via smooth deformations in
the two-dimensional plane and finite set of Reidemeister moves,
shown in Fig.\ref{rei}. \bfig \bc
\includegraphics[scale=0.45]{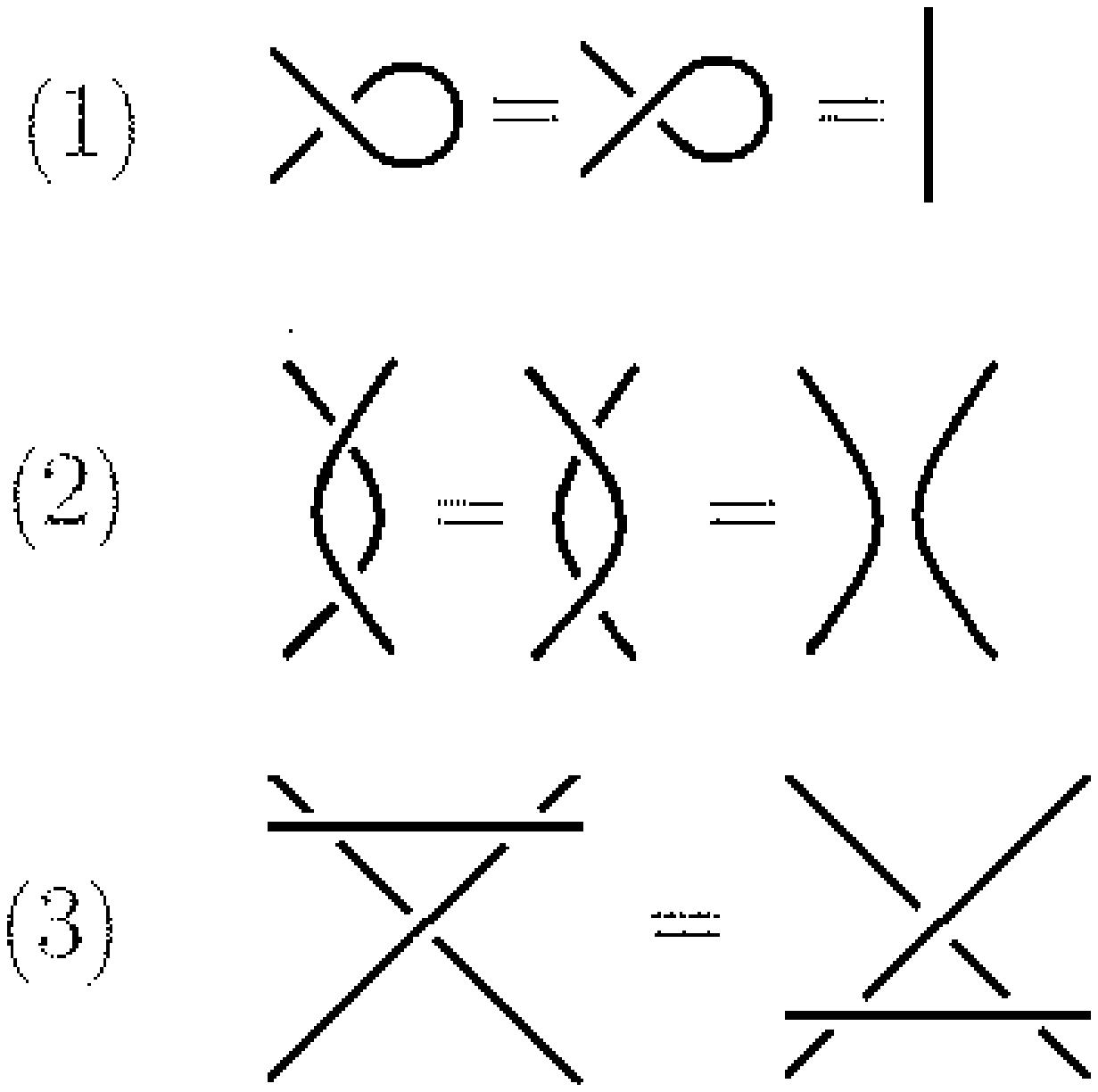}  \caption{\label{rei}} \ec
\efig }
 The quantities that depend only on equivalence
classes of ambient isotopy, or equivalently, do not vary under
smooth two-dimensional deformations and  Reidemeister moves are
referred to as the \textit{ambient isotopy invariants}.
\subsection{Temporal framing. Regular isotopy invariants. \label{framing} }
Despite the topological nature of CST, the expectation values of
Wilson loops in this theory are not ambient isotopy invariants
\cite{Gu1}-\cite{Gu12}.  This feature of CST is closely related to
the self-linking problem and the \textit{framing} procedure. In
CST the vevs of Wilson loops $<W(c)>$ are ill-defined quantities,
and to define these quantities properly one can introduce a
second, auxiliary contour $c^{'}$, called framing of $c$, which
can be considered as a slight displacement of contour $c$ along a
normal to $c$ vector field \cite{WCS}. More concretely, the word
"slight" means that we replace the contour $c$ representing the
knot by a narrow-width rope bounded by $c$ and $c^{'}$. Two
different choices of framing contour for straight line are
presented in the Fig.\ref{kn}($a$). When auxiliary contour is
chosen, then it makes sense to calculate the linking number
between contours $c$ and $c^{'}$ and its non-Abelian
generalizations - the normalized vevs of product of two Wilson
loops in CST: \be\label{nor} <W(c)> \, \stackrel{\textrm{def}}{=}
\, <:W(c):\,:W(c^{'}):>\ee The normalization means that in
perturbation theory only propagators with different ends attached
to different contours $c$ and $c^{'}$ are taken into account.
Therefore, there are no collapsible propagators, and the quantity
(\ref{nor}) turns out to be well defined.

 Roughly speaking, the CST describes the properties of
knots that are made not of a "rope" but of a "ribbon" bounded by
$c$ and $c^{'}$. In this case an additional parameter comes into
the game. Namely, a "rope" can be replaced by a "ribbon" in many
different ways with different number of twists $n$ of the "ribbon"
Fig.\ref{kn}($a$). The answer in CST depends only on this $n$. To
fix the number of the ribbon twists, different framing fixing are
used. In the temporal gauge there exists a natural choice for
framing fixing. Namely, one can fix two dimensional plane $
\mathbb{R}^2 \subset \mathbb{R}^3$ and then only ribbons which
have non-perpendicular to this plane tangent planes at every point
are considered. We call this procedure - \textit{temporal
framing.} The examples of knots in the temporal framing are
represented in the Fig.\ref{kn}($b$) , here the fixed
two-dimensional plane is the plane of the page.\\
 \vspace{5mm}
\bfig \bc
\includegraphics[scale=0.45]{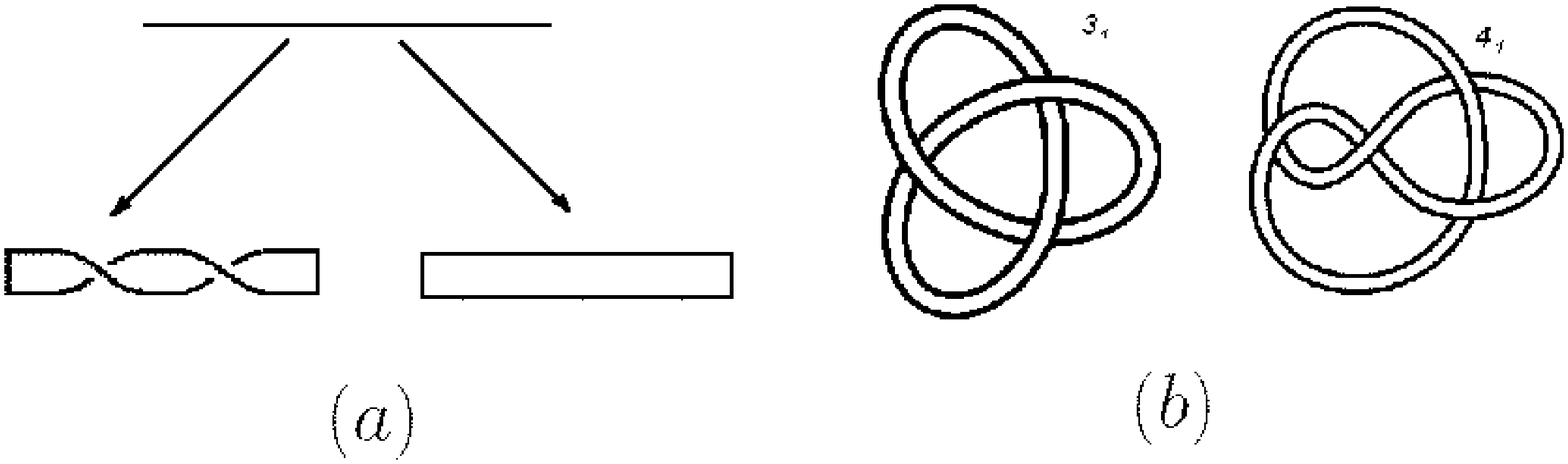}  \caption{\label{kn}} \ec
\efig

The "Wilson-loop" calculations in CST depend on the definition
 of the framing. In what follows we assume that all quantities are
represented in the temporal framing.

 Now, it is obvious that in the temporal
framing the first Reidemeister move from Fig.\ref{rei} is not
respected, because the ribbon with loop on it in the temporal
framing is not equivalent to a straight ribbon, but to a ribbon
with additional twist, see Fig.\ref{fra}\\
 \bfig \bc
\includegraphics[scale=0.8]{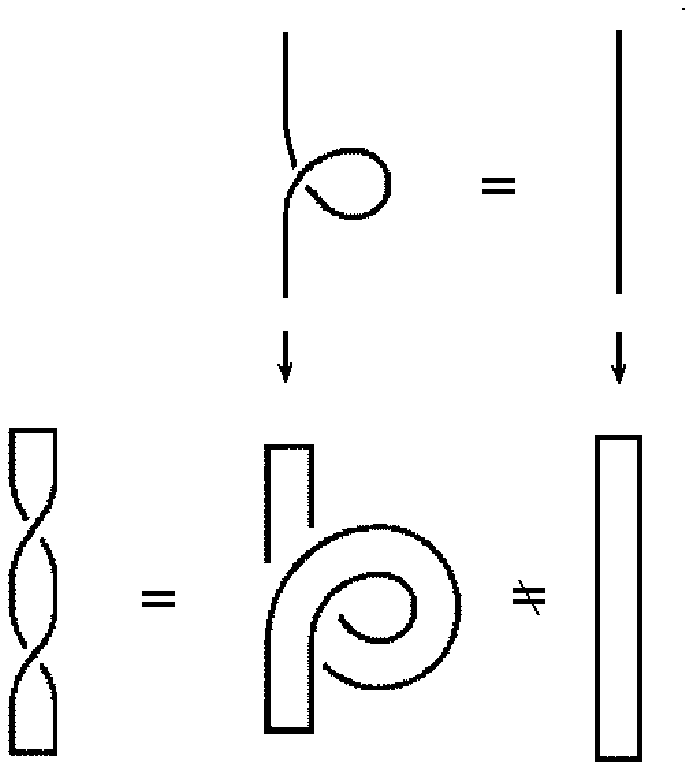}  \caption{\label{fra}} \ec
\efig

 The equivalence of knot projections defined only
by the second  and the third Reidemeister moves form Fig.\ref{rei}
is usually referred to as \textit{regular isotopy invariance}, and
quantities that does not change under these moves are the so
called \textit{regular isotopy invariants}. In this way, we see,
that the presence of the framing result in the fact that the vevs
of Wilson loops are not ambient but only the regular invariants of
knots.\\
\indent It was shown in the Witten's work \cite{WCS} that the
shift of the framing contour by $n$ twists results in a simple
multiple for the vev of a Wilson loop:
$$
<W(c)>\, \rightarrow\, \alpha^{n}\,<W(c)>
$$
This is a crucial observation that enables the CST to work. It
means that although we must pick additional framing contour to
make the vev well defined, we do not lose the information since we
know how this vev changes a under change of framing. Explicit
perturbative computations performed in \cite{Gu1} show that the
first terms of $2\,\pi/k_{1}$ expansion of $\alpha$ are:
$$
\alpha=1+\dfrac{2\,\pi}{k_{1}}\,
\varrho\,(\Omega_{2})+\dfrac{1}{2} \,
\left(\dfrac{2\,\pi}{k_{1}}\, \varrho\,(\Omega_{2})\right)^2 +
O\left( \dfrac{1}{k^3_{1}}
 \right)
$$
where $k_{1}$ is  renormalization  of the bare constant $k$,
$\Omega_{2}$ is the value of quadratic Casimir of the gauge Lie
algebra and $\varrho$ is the representation carried by the Wilson
loop. We see that the results of the calculations are in agreement
with the following expression: \be \label{twist}
\alpha=q^{\,\varrho\left( \Omega_{2}\right)} , \ \ \ q =\exp\left(
\dfrac{2\,\pi\,i}{k_{1}}\right)\ee The difference between $k_{1}$
and $k$ is somewhat controversial point, usually one assumes that
$k_{1}= k+ \varrho_{Ad}( \Omega_{2} )$ (here $\varrho_{Ad}(
\Omega_{2} )$ stands for the value of quadratic Casimir in the
adjoint representation of the gauge group) like in $2d$ WZNW
theory \cite{GMMOS}. For our guessing this is not very essential:
in the following we use directly parameter $q$ irrespective of its
exact dependence of the bare $k$.

Therefore, the CST generalizes the first Reidemeister move form
Fig.\ref{rei} and the rest two remains unchanged. Finally, we can
infer that the vevs of Wilson loops have the transformation
properties summarized in the Fig.\ref{reid}\\
 \vspace{7mm}
 \bfig \bc
\includegraphics[scale=0.5]{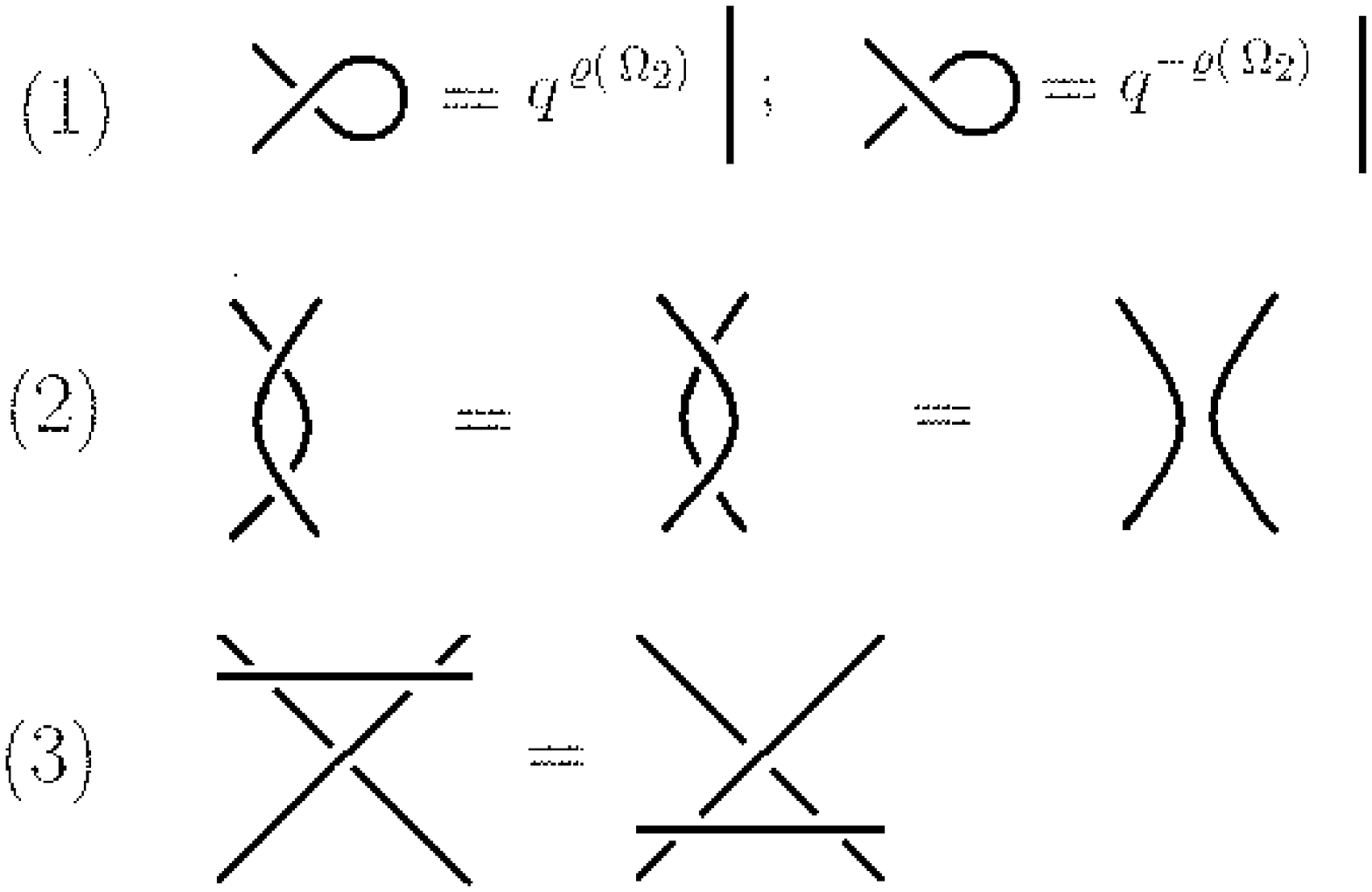}  \caption{\label{reid}} \ec
\efig
\subsection{The fundamental equations}
Let us analyze the restrictions imposed by generalized
Reidemeister moves Fig.\ref{reid} on the crossing and the turning
point operators. First of all, we note that using the relations
(\ref{tur})-(\ref{crossmin}) we can express the
 crossing operators through ${\cal R}$ and $\overline{{\cal R}}$,
similarly, the turning point operators are expressed trough ${\cal
M}$ and $\overline{{\cal M}}$. Therefore, the resulting
restrictions on the crossing and turning point operators can be
reduced to equations on ${\cal R},\,\overline{{\cal R}}, {\cal M}$
and $\overline{\cal M}$.  In terms of ${\cal R}$ and
$\overline{{\cal R}},$ the second Reidemeister relation in
Fig.\ref{reid3} can be expressed as:
 \be  \label{fir}
\overline{{\cal R}}^{\,a\,b}_{\,c\,d}\,{\cal
R}^{\,c\,d}_{\,e\,f}={\cal R}^{\,a\,b}_{\,c\,d}\,\overline{{\cal
R}}^{\,c\,d}_{\,e\,f}=\delta^{a}_{e}\,\delta^{b}_{f}, \ \ \
\textrm{or} \ \ \ \ \overline{{\cal R}}\,= {\cal R}^{-1}    \ee
The first relation in Fig.\ref{reid3} means that: \be \label{QR}
{\cal R}^{\,a\,e}_{\,b\,c}\, {\cal M}^{\,c}_{\,d}\,
\overline{{\cal M}}^{\,d}_{\,e}\,= q^{\,\varrho(\Omega_{2})}
\,\delta^{a}_{b}\, ,\ \ \ \ \ \overline{{\cal
R}}^{\,a\,e}_{\,b\,c}\, {\cal M}^{\,c}_{\,d} \,\overline{{\cal
M}}^{\,d}_{,e}\,= q^{- \varrho\,(\Omega_{2})} \,\delta^{a}_{b} \ee
or, equivalently: \be \tr_{2}( {\cal R}^{\,\pm\,1} \,
1\otimes{\cal Q} )\,= q^{ \, \pm \,\varrho\,(\Omega_{2}) } \ee
 where
$\tr_{2}$ denotes the trace over the second space in the tensor
product and we denote: \be \label{qmm} {\cal Q}={\cal M}\,
\overline{{\cal M}}. \ee
%Note
%that the generalized Reidemeister moves define the equations
%(\ref{QR}) that relate the product of two turning point operators
%${\cal Q}={\cal M}\, \overline{{\cal M}}$ with crossing operators
%${\cal R}$ and $\overline{{\cal R}}$. Therefore, to construct the
%turning points operators additional relation between ${\cal M}$
%and $\overline{\cal M}$ is necessary. For the sake of symmetry we
%assume that: \be \label{addrel} {\cal M}=\overline{{\cal M }}, \ \
%\ \textrm{and} \ \ \ {\cal M}^{2}= {\cal Q} \ee
 In terms of the "down-directed" crossings corresponding to the operator ${\cal R}$  in  Fig.\ref{reid3}(3)  third Reidemeister move
 implies that ${\cal R}$ satisfies the following trilinear equation:
 \be
 \label{thir}
 {\cal R}^{\,a\,b}_{\,d\,e}{\cal R}^{\,e\,c}_{\,f\,g}{\cal
 R}^{\,d\,f}_{\,k\,m}=
 {\cal R}^{\,b\,c}_{\,d\,e}{\cal R}^{\,a\,d}_{\,k\,m}{\cal
 R}^{\,g\,e}_{\,m\,g}
 \ee
It is easy to check that all other possible
 orientations of the crossings in the third Reidemeister relation
 Fig.\ref{reid}\,(3) can be reduced to (\ref{thir}) by
 transformations  (\ref{crosspl}) and (\ref{crossmin}). For this reason only (\ref{thir}) should be
 considered.
 Introducing the notations ${\cal R}_{1 2}={\cal R} \otimes 1,\ \ {\cal R}_{2 3}= 1 \otimes {\cal R} $
  for the operators, acting in the triple product $V\otimes V\otimes V$, we can rewrite this equation in the form:
 \be
 \label{QYBE}
{\cal R}_{1 2}\,{\cal R}_{2 3}\,{\cal R}_{1 2}={\cal R}_{2
3}\,{\cal R}_{1 2}\,{\cal R}_{2 3}
 \ee
 This is the famous quantum
 Yang-Baxter equation (QYBE), arising in a number of physical and mathematical
 topics,
 from exactly solvable statistical models \cite{Bax} to quantum
 groups \cite{MV1}-\cite{qalg1}.
% Let us summarize the main results of this section.
 Therefore,
 %demonstrated that
 to describe the properties of vevs of Wilson loops in the CST the crossing
 and the turning point operators must satisfy two
 \textit{fundamental} equations:
 \be
 \label{fundeq}
 \begin{array}{|c|}
 \hline
 \\
  \tr_{2}\left(  {\cal R}^{\pm 1}\ \
 1\otimes {\cal  Q } \right)= q^{\,\pm\,
 \Omega_{2}} \\
 \\
 {\cal R}_{1 2}\,{\cal R}_{2 3}\,{\cal R}_{1 2}={\cal R}_{2
 3}\,{\cal R}_{1 2}\,{\cal R}_{2 3}\\
 \\
 \hline
 \end{array}
 \ee
 Solution of these equations is a pair $({\cal R}, {\cal Q})$ of
 operators acting in appropriate tensor products of gauge group representations.
 With such a solution at hand, one can construct all crossing
 operators with the help of relations (\ref{crosspl}), (\ref{crossmin}) and (\ref{fir}). On the other hand side the pair $({\cal R}, {\cal
 Q})$
 satisfying (\ref{fundeq}) does not define the turning point
 operators but only the product of two operators (\ref{qmm}). There is
 no additional relation between ${\cal M}$ and $\overline{{\cal M}}$
 imposed by topological moves and arbitrary choice of ${\cal M}$ and $\overline{{\cal M}}$
 satisfying (\ref{qmm}) is possible. For instance, one can consider the following choice:
 $$
{\cal M} ={\cal Q}, \ \ \ \overline{{\cal M}}=1
 $$
 what means that the turning point operators ${\cal M}_{2}$ and ${\cal M}_{4}$
  are identity operators due to (\ref{turrel })  and do not give contribution to vev of Wilson loops and
therefore, the corresponding
 turning points are not taken into account. In what follows we,
 however, assume the following relation:
 \be \label{adrel} {\cal M} =\overline{{\cal M}},\ \ \ \ {\cal M}^{2}=\,{\cal Q} \ee
 This choice of additional relation is for the sake of symmetry
 as it makes all the four turning point operators and turning points
 equivalent. When added to (\ref{crosspl}), (\ref{crossmin}), (\ref{turrel }), the relation (\ref{adrel}) allows to
 construct all crossing and turning point operators from given
 solution of fundamental equations.\\
 \indent

\indent
 \bfig \bc
\includegraphics[scale=0.4]{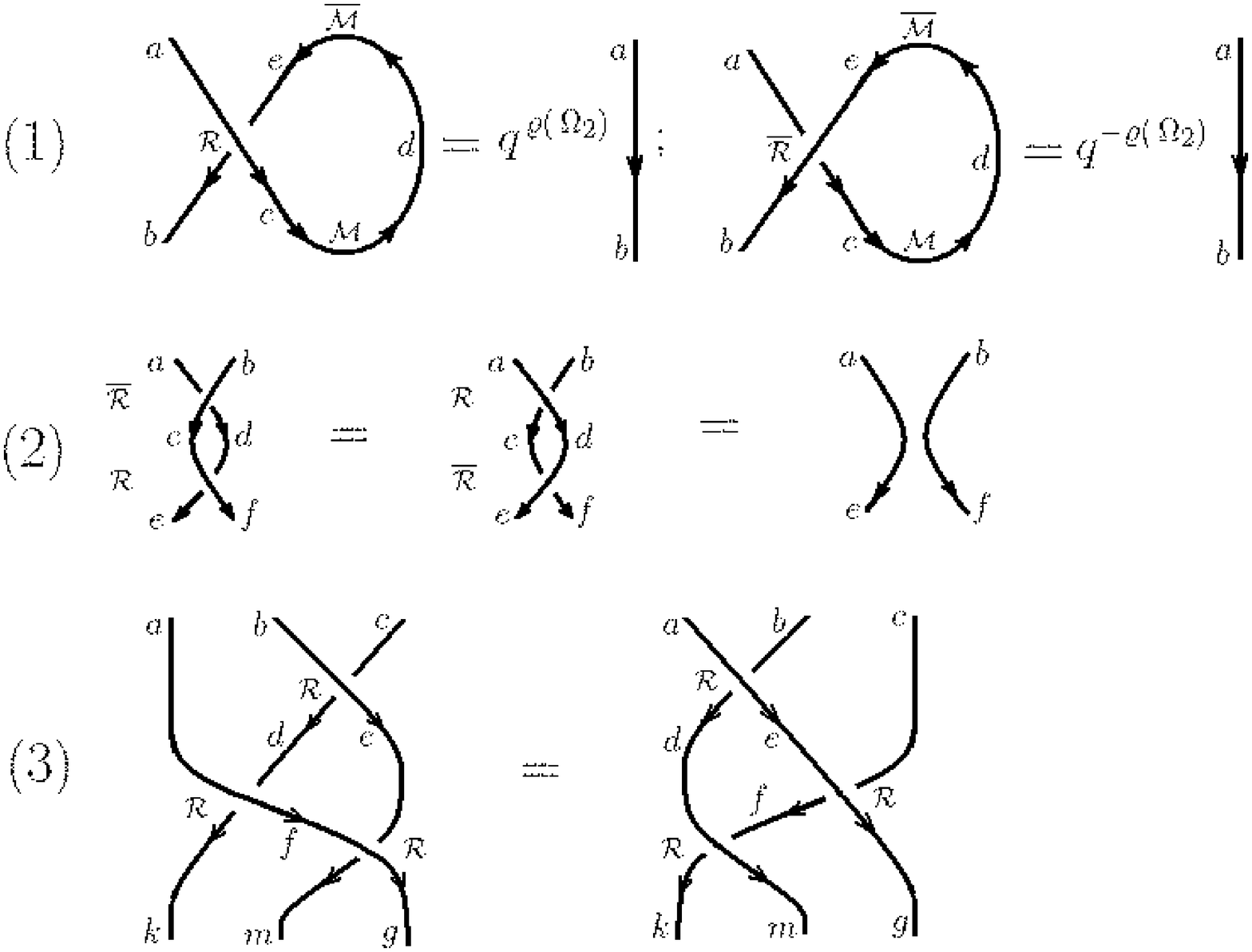}  \caption{\label{reid3}} \ec
\efig
\section{Universal quantum $R$-matrix \label{Sol}}
As it was shown in the previous paper \cite{Sm}, the use of the
naive propagator (\ref{prop}) leads to the crossing operator of
the form: \be \label{naiveR}
 R_{\varrho}=\sum_{m=0}^{\infty} \dfrac{h^{m}}{m!}
\{T^{a_{1}}_{\varrho}\,T^{a_{2}}_{\varrho}...\,T^{a_{m}}_{\varrho}\}\otimes
\{T^{a_{1}}_{\varrho}\,T^{a_{2}}_{\varrho}...\,T^{a_{m}}_{\varrho}\},\
\ \ \
\{T^{a_{1}}_{\varrho}\,T^{a_{2}}_{\varrho}...\,T^{a_{m}}_{\varrho}\}=\dfrac{1}{m!}
\sum\limits_{\sigma\in {\cal S}_{m}}
T^{a_{\sigma(1)}}_{\varrho}\,T^{a_{\sigma(2)}}_{\varrho}...\,T^{a_{\sigma(m)}}_{\varrho}
\ee where ${\cal S}_{m}$ is the symmetric group of $m$-elements,
$\varrho\,$ is a representation carried by a Wilson loop and
$h=2\pi i/k_{1}$. The universal, representation independent, form
of (\ref{naiveR}) is: \be \label{UnivNaiveR}
 {\cal
R}=\sum_{k=0}^{\infty} \dfrac{h^{m}}{m!}
\{T^{a_{1}}\,T^{a_{2}}...\,T^{a_{m}}\}\otimes
\{T^{a_{1}}\,T^{a_{2}}...\,T^{a_{m}}\} \in  U(g)\otimes U(g) \ee
where $U(g)$ is the universal enveloping algebra of $g$.
Unfortunately, such  crossing operator (\ref{UnivNaiveR}) does not
satisfy the QYBE (\ref{QYBE}), what indicates that mote accurate
regularization of propagator is necessary. Obviously, such a
regularization of the naive answer should deform
 (\ref{UnivNaiveR}) somehow. In this section we make an obvious suggestion
 that this deformation of (\ref{UnivNaiveR}) is given by the well known
 \textit{quantum universal $R$-matrix} that seems to be in accordance with \cite{Ka4, K1, Rm1}. The quantum universal
 $R$-matrix is not the element of $U(g)\otimes U(g)$ anymore but
 the element of $U_{h}(g)\otimes U_{h}(g)$, where the algebra
 $U_{h}(g)$ is the $h$-deformed, or the quantized universal enveloping
 algebra of $g$. In the subsections \ref{sub1}-\ref{sub3} we
 remind explicitly definition of $U(g)$ and $U_{h}(g)$. In the subsection
 \ref{sub4}
 we give the explicit expression for the universal quantum $R$-matrix
 ${\cal R}\in U_{h}(g)\otimes U_{h}(g)$, then define the element ${\cal Q}\in
 U_{h}(g)$ and prove that these elements satisfy the fundamental
 equations (\ref{fundeq}).

\vspace{7mm}
\subsection{Universal enveloping algebras \label{sub1}}
Let $g$ be a simple Lie algebra of rank $r$ with the root system
$\Phi$. Denote by $\Phi^{+}$ the subset of positive roots and by
$n=| \Phi^{+} |$ be the number of positive roots. Fix the set of
simple roots $\{\alpha_{1},..., \alpha_{r}\}= \Delta$,\,
%and denoteby $a_{i j}=<\alpha_{i},\,\alpha_{j}>$ the corresponding Cartan
%matrix.
 and let $\{ \alpha_{1}^{\vee},...,\alpha_{r}^{\vee}: \alpha
\in \Delta \}$ be the set of elements
dual to the roots elements:\\
\be (\alpha^{\vee},\beta)\,=\delta_{\alpha\beta}\,=0\ \
\textrm{if} \ \ \alpha\neq\beta \ \ \textrm{and} \ \ 1 \ \
\textrm{otherwise} \ee Here (\,,\,) is the Killing form reduced to
the Cartan subalgebra. Let $\{e_{\alpha}, f_{\alpha}, h_{\beta}:
\alpha\in \Phi^{+}, \beta\in\Delta \}$ be a Cartan-Weil basis of
$g$. The Cartan matrix of the algebra is defined as
$a_{\alpha\beta}=2\,(\alpha\,\beta)/(\alpha\,\alpha)$. For
arbitrary element $\gamma$ from the Cartan subalgebra, we define
the corresponding elements of the algebra as follows: \be
h_{\gamma}\,= \,\sum\limits_{\alpha\in \Delta}
h_{\alpha}\,(\alpha^{\vee},\gamma \,) \ee In particular, we will
need the elements $h_{\beta}, \ \ \beta\in \Phi^{+}$, and the
element $h_{\rho}$ where $\rho$ is the half-sum of all positive
roots: \be \rho=\dfrac{1}{2}\sum\limits_{\alpha\in \Phi^{+}}\alpha
\ee
% \indent For
%every Lie algebra $g$ there exist the Cartan-Weil elements (not
%uniquely defined) $ \{e_{\alpha}, f_{\alpha}, h_{\beta}: \alpha\in
%\Phi^{+},\, \beta~\in~\Delta \}$ which form a basis of g. In the
%Cartan-Weil basis the commutation relations read: \be
%[h_{\alpha},h_{\beta}]=0,\ \
%[h_{\alpha},e_{\beta}]=<\alpha\,\beta> e_{\beta},\ \
%[h_{\alpha},f_{\beta}]=-<\alpha\,\beta> f_{\beta}\ee \be
%[e_{\alpha}\,f_{\beta}]=0 \ \ \ \textrm{if}\ \ \ \alpha\neq \beta,
%\ \ \ 2 h_{\alpha}\ \ \  \textrm{if}\ \ \ \alpha=\beta\ee \be
%[e_{\alpha},e_{\beta}]=[f_{\alpha},f_{\beta}]=0 \ \ \ \textrm{if}
%\ \ \ \alpha+\beta \notin \Phi \ee
 The universal enveloping
algebra of $g$ is the quotient of the tensor algebra $ T(g) $ by
the ideal generated by the  elements $ x\otimes y-y\otimes
x=[x,y]$: \be U(g)=T(g)/I,\ \ \ \ \
T(g)=\bigoplus\limits_{m=0}^{\infty} g^{\otimes m},\ \ \ \ \
I=\{x\otimes y-y\otimes x=[x,y]\} \ee According to the Serre's
theorem this algebra can be defined as the algebra generated by
$3\,r$ elements $\{e_{\alpha}, f_{\alpha}, h_{\alpha}:\alpha\in
\Delta\}$ with the following relations\footnote{ Note, that our
choice of Cartan elements $h_{\alpha}$ is different from the
conventional one by the factor $(\alpha,\alpha)/2$.}: \be
\label{serr1} [\,h_{\alpha},h_{\beta}\,]=0,\,
[\,h_{\alpha},e_{\beta}\,]\,=\,(\alpha, \beta)\, e_{\beta},\ \
[\,h_{\alpha},f_{\beta}\,]\,=\,-(\alpha, \beta)\, f_{\beta}\ee\be\
\label{serr2} [\,e_{\alpha},f_{\beta}\,]\,=
\,\delta_{\alpha\beta}\, h_{\alpha} \ee \be\label{serr3}
(\textrm{ad}\, e_{\alpha})^{1-\,a_{\alpha\beta} } e_{\beta}\equiv
\sum \limits_{m=0}^{1-a_{\alpha\beta}}(-)^{m}\left[
1-a_{\alpha\beta} \atop m \right]
e_{\alpha}^{1-\,a_{\alpha\beta}-m}\,e_{\beta}\,e_{\alpha}^{m} =0,\
\ \alpha\neq\beta \ee\ \be \label{serr4}  (\textrm{ad}\,
f_{\alpha})^{1-\,a_{\alpha\beta} } f_{\beta}\equiv \sum
\limits_{m=0}^{1-\,a_{\alpha\beta}}(-)^{m}\left[ 1-a_{\alpha\beta}
\atop m \right]
f_{\alpha}^{1-\,a_{\alpha\beta}-m}\,f_{\beta}\,f_{\alpha}^{m} =0,\
\ \alpha\neq\beta\ee The possible choice of a basis in $U(g)$ is
given by the following Poincar\'e-Birchoff theorem:\\
\textit{The elements
$$
e_{\alpha_{1}}^{p_{1}}...e_{\alpha_{n}}^{p_{n}}\,h_{\beta_{1}}^{t_{1}}...h_{\beta_{r}}^{t_{r}}\,f_{\alpha_{n}}^{s_{1}}...f_{\alpha_{1}}^{s_{n}},\
\ \ \alpha_{i}\,\in \Phi^{+},\ \ \ \beta_{i}\in \Delta, \ \
p_{i},t_{i},s_{i}\in \mathbb{N}_{0}
$$
form a basis of the universal enveloping algebra $U(g)$.}
\vspace{5mm}
\subsection{Quantized universal enveloping algebras \label{sub2}}
 Let $h$ be indeterminate number, then the quantized universal enveloping
algebra $U_{h}(g)$ is the algebra with $1$ over
$\mathbb{C}[\,q\,]$, $q=e^h$ generated by $3\,r$ elements
$\{h_{\alpha},\, E_{\alpha},\, F_{\alpha},\, \alpha\in \Delta \}$
subjected to the relations: \be \label{qserr1}
[\,h_{\alpha},\,h_{\beta}\,]\,=\,0,\ \
[\,h_{\alpha},\,E_{\beta}\,]\,=\,\, (\alpha,\beta)\, E_{\beta},\ \
[\,h_{\alpha},\,F_{\beta}\,]\,=\,(\alpha, \beta)\, F_{\beta} \ee

\be \label{qserr2}[\,E_{\alpha},\,F_{\beta}\,]\,=\,\delta_{\alpha
\beta}\,
\dfrac{q_{\alpha}^{h_{\alpha}}-q_{\alpha}^{-h_{\alpha}}}{q_{\alpha}-q_{\alpha}^{-1}}
\ee \be \label{qserr3}
\sum\limits_{m=0}^{1-\,a_{\alpha\beta}}(-)^{m}\left[1-\,a_{\alpha\beta}\atop
m \right]_{q_{\alpha}}\,E_{\alpha}^{1-\,a_{\alpha\beta}-m}
\,E_{\beta}\,E_{\alpha}^{m}=0, \ \ \alpha\neq\beta \ee \be
\label{qserr4}
\sum\limits_{m=0}^{1-\,a_{\alpha\beta}}(-)^{m}\left[1-\,a_{\alpha\beta}\atop
m \right]_{q_{\alpha}}\,F_{\alpha}^{1-\,a_{\alpha\beta}-m}
\,F_{\beta}\,F_{\alpha}^{m}=0, \ \ i\neq j \ee where \be \left[
n\atop m\right]_{q}=\dfrac{[n]_{q}!}{[m]_{q}!\,[n-m]_{q}!},\ \ \
[n]_{q}!=\prod\limits_{k=1}^{n}[k]_{q},\ \ \
[n]_{q}=\dfrac{q^{n}-q^{-n}}{q-q^{-1}},\ \  \textrm{and}\ \
q_{\beta}=q^{\,(\alpha,\alpha)/2} \ee Note, that the definition of
$U_{h}(g)$ is similar to the definition of $U(g)$ by the Serre
relations (\ref{serr1})-(\ref{serr4}). The relations
(\ref{qserr1}) are formally equivalent to the relations
(\ref{serr1}), and the relations (\ref{qserr2})-(\ref{qserr4}) are
some $h$-deformations of the respective relations
(\ref{serr2})-(\ref{serr4}). \vspace{5mm}
\subsection{Braid groups and root elements\label{sub3}}
If $a_{\alpha\beta}$ is the Cartan matrix of a Lie algebra $g$,
then the numbers $a_{\alpha \beta}a_{\beta\alpha}$ can take values
$0,1,2$ or $3$. Let $m_{\alpha \beta}$ be equal to $2,3,4$ or $6$
when $a_{\alpha\beta}a_{\beta\alpha}$ is equal to $0,1,2$ or $3$
respectively. The Weil group $W_g$ is defined by reflections
$\{\sigma_{\alpha}:\, \alpha\in \Delta \}$ corresponding to the
simple roots of $g$ satisfying the following relations: \be
\sigma_{\alpha}^2=1,\ \ \ \
\underbrace{\sigma_{\alpha}\sigma_{\beta}\sigma_{\alpha}\sigma_{\beta}...}_{m_{\alpha
\beta}\, \textrm{times}}
=\underbrace{\sigma_{\beta}\sigma_{\alpha}\sigma_{\beta}\sigma_{\alpha}...}_{m_{\alpha
\beta}\, \textrm{times}} \,  ,\ \ \alpha\neq \beta   \ee  Let
$\sigma_{0}$ be the longest element of the Weil group i.e. the
element that has the longest length of its reduced decomposition
$\sigma_{0}=\sigma_{\alpha_{i_{1}}}\sigma_{\alpha_{i_{2}}}...\sigma_{\alpha_{i_{n}}},
\ \ \alpha_{i_{k}}\in \Delta$ (length  of this element is equal to
the number of positive roots $n$ for every Lie algebra). An
important property of $\sigma_{0}$ is that the sequence: \be
\label{roots} \beta_1=\alpha_{i_{1}},\ \
\beta_{2}=\sigma_{\alpha_{i_{1}}} \alpha_{i_{2}},\ \  ... ,\ \
\beta_{n}=\sigma_{\alpha_{i_{1}}}\sigma_{\alpha_{i_{2}}}...\sigma_{\alpha_{i_{n-1}}}
\alpha_{i_{n} } \ee exhaust all positive roots of $g$ and defines
the ordering $\beta_{1}<\beta_{2}<...<\beta_{n}$ on the set
$\Phi^{+}$. Therefore, the Weil group and a fixed reduced
decomposition of a longest element $\sigma_{0}$ allow  to define
all root elements of $U(g)$ starting from simple roots. \\
\indent In the definition of $U_{h}(g)$ enter only elements
$h_{\alpha}, E_{\alpha}$ and $F_{\alpha}$ corresponding to simple
roots. To define the elements of $U_{h}(g)$ corresponding to the
non-simple roots we need a notion of \textit{quantum Weil group}
or \textit{Artin braid group}.
\\
\indent The Artin braid group $B_{g}$ associated with $g$ is the
group generated by elements $\{b_{\alpha}: \alpha\in \Delta \}$
with the following relations: \be
\underbrace{b_{\alpha}b_{\beta}b_{\alpha}b_{\beta}...}_{m_{\alpha
\beta}\, \textrm{times}}
=\underbrace{b_{\beta}b_{\alpha}b_{\beta}b_{\alpha}...}_{m_{\alpha
\beta}\, \textrm{times}} \, ,\ \ \alpha\neq \beta \ee The braid
group $B_{g}$ acts by automorphisms on $U_{h}(g)$ \cite{qalg1}:
\be b_{\alpha}(h_{\beta})=h_{\alpha}-a_{\beta\alpha}h_{\alpha},\ \
b_{\alpha}(E_{\alpha})=-F_{\alpha}q_{\alpha}^{h_{\alpha}},\ \
b_{\alpha}(F_{\alpha})=-E_{\alpha}q_{\alpha}^{-h_{\alpha}} \\
b_{\alpha}(E_{\alpha})=\sum\limits_{m=0}^{-a_{\alpha \beta}}\,
\dfrac{(-)^{m-a_{\alpha
\beta}}}{[-a_{\alpha \beta}-m]_{q_{\alpha}}![m]_{q_{\alpha}}!  }\,q_{\alpha}^{-m} (E_{\alpha})^{-a_{\alpha \beta}-m}E_{\beta} (E_{\alpha})^{m} \\
b_{\alpha}(F_{\beta})=\sum\limits_{m=0}^{-a_{\alpha \beta}}
\dfrac{(-)^{m-a_{\alpha \beta}}}{[-a_{\alpha
\beta}-m]_{q_{\alpha}}![m]_{q_{\alpha}}!}\,q_{\alpha}^{m}
(F_{\alpha})^{m}F_{\beta} (F_{\alpha})^{-a_{\alpha \beta}-m}\ee
Let us define the roots of $g$ as in (\ref{roots}), then we define
the root elements of $U_{h}(g)$ as follows: \be
E_{\beta_{k}}=b_{\alpha_{i_{1}}}b_{\alpha_{i_{2}}}...b_{\alpha_{i_{k-1}}}E_{\alpha_{i_{k}}}
\ \ \textrm{and} \ \ \
F_{\beta_{k}}=b_{\alpha_{i_{1}}}b_{\alpha_{i_{2}}}...b_{\alpha_{i_{k-1}}}F_{\alpha_{i_{k}}},\
\ \ k=1,...,n \ee The following analog of Poincar\'e-Birchoff
theorem gives a basis in
$U_{h}(g)$:\\
\\
\textit{The elements}
$$
F_{\alpha_{1}}^{p_{1}}...F_{\alpha_{n}}^{p_{n}}\,h_{\beta_{1}}^{t_{1}}...h_{\beta_{r}}^{t_{r}}\,E_{\alpha_{n}}^{s_{1}}...E_{\alpha_{1}}^{s_{n}}\,,\
\ \  \alpha_{i} \in \Phi^{+},\ \ \beta_{i}\in \Delta,\ \
p_{i},\,t_{i},\,s_{i}\in \mathbb{N}_{0}
$$
\textit{form a basis of} $U_{h}(g)$.\\
\\
 For a proof see \cite{qalg1}.
\vspace{7mm}
\subsection{Universal quantum $R$-matrix \label{sub4}}
Let $E_{\beta}$ and $F_{\beta}$, $ \beta\in \Phi^{+} $ are the
root elements of $U_{h}(g)$ then the universal quantum $R$-matrix
is defined as an element of $U_{h}(g)\otimes U_{h}(g)$ by the
following explicit expression: \be\label{quR}
\begin{array}{|l|}
\hline \\ {\cal R}= {\hat P} \, q^{\,\sum\limits_{\alpha\in \Delta
}h_{\alpha}\otimes\, h_{\alpha^{\vee}}} \,
\prod\limits_{\beta\in\,\Phi^{+}}^{\rightarrow}
\exp_{q_{\beta}}\left( (q_{\beta}-q_{\beta}^{-1}) E_{\beta}\otimes
F_{\beta} \right) \\  \\ \hline \end{array} \ee Here ${\hat P}$ is
the permutation operator: ${\hat P}\, a\otimes b=b\otimes a\,
{\hat P}$  and the arrow above the product implies that the
factors appear in the order $\beta_{n},\beta_{n-1},...,\beta_{1}$
defined by a reduced decomposition of longest elements of Weil
group (\ref{roots}). The q-exponent is defined as:
$$
\exp_{q}(A)=\sum\limits_{m=0}^{\infty}
\dfrac{A^{m}}{[m]_{q}!}\,q^{m(m-1)/2}
$$
The crucial property of the universal quantum $R$-matrix is that
this matrix represents suitable regularization of the crossing
operator
(\ref{UnivNaiveR}). More precisely:\\
\\
\indent\textbf{Proposition:} \textit{The pair $\{ {\cal R}\in
U_{h}(g)\otimes U_{h}(g),\,{\cal Q}\in U_{h}(g) \}$, where ${\cal
R}$ is the universal quantum $R$-matrix (\ref{quR}) and ${ \cal
Q}$ is defined by the following explicit expression: \be \label{Q}
{\cal Q}= q^{\,h_{\rho}},\ \ \ \rho =\dfrac{1}{2}
\sum\limits_{\alpha\,\in\, \Phi^{+}}\alpha, \ \ \
h_{\rho}=\sum\limits_{\alpha\in \,\Phi^{+}}\,h_{\alpha} \ee
satisfies the fundamental equations,
(\ref{fundeq}).}\\
\\
\indent \textbf{Proof:} The proof of the quantum Yang-Baxter
equation (the second fundamental equation) can be found in any
textbook on quantum group theory,see for example \cite{MV2} or
\cite{qalg1}. The proof of the first fundamental equation in
(\ref{fundeq}) is a relatively long but direct computation:
\be\nonumber \tr_{2}({\cal R}\, 1\otimes {\cal Q})= \tr_{2}
\left\{{\hat P} \exp( h \sum\limits_{\alpha\in\,\Delta}
h_{\alpha}\otimes h_{\alpha^{\vee}} )
\prod\limits_{\beta\in\,\Phi^{+}}^{\rightarrow}
\exp_{q_{\beta}}\left( (q_{\beta}-q_{\beta}^{-1})\,
E_{\beta}\otimes
F_{\beta} \right) 1\otimes q^{h_{\rho}}\right\}= \\
\label{step1} =\tr_{2}\,\left\{{\hat P}
\sum\limits_{k=0}^{\infty}\, \dfrac{h^k}{k!}
\sum\limits_{\alpha_{1}...\,\alpha_{k}\in
\Delta}\,\sum\limits_{m_{\beta},\,\beta\in\,\Phi^{+} }^{\infty}
\,\prod\limits_{\eta\in\,\Phi^{+} }
\dfrac{(q_{\eta}-q_{\eta}^{-1})^{m_{\eta}}}{[m_{\eta}]_{q_{\eta}}!}\,
q^{m_{\eta}(m_{\eta}-1)/2}_{\eta}
\prod\limits_{p=1}^{k}h_{\alpha_{p}} \prod\limits_{\gamma \in
\,\Phi^{+} }^{\rightarrow} E_{\gamma}^{m_{\gamma}} \otimes
\prod\limits_{s=1}^{k}h_{\alpha_{s}^{\vee}} \prod\limits_{\delta
\in\,\Phi^{+}
}^{\rightarrow} F_{\delta}^{m_{\delta}}\,q^{H_{\rho}}\right\}\\
\label{step2}
 =\sum\limits_{k=0}^{\infty}\,
\dfrac{h^k}{k!} \sum\limits_{\alpha_{1}...\,\alpha_{k}\in
\,\Delta}\,\sum\limits_{m_{\beta}, \,\beta\in\,\Phi^{+}}
\prod\limits_{\eta \in \,\Phi^{+} }
\dfrac{(q_{\eta}-q_{\eta}^{-1})^{m_{\eta}}}{[m_{\eta}]_{q_{\eta}}!}\,
q^{m_{\eta}(m_{\eta}-1)/2}_{\eta}\,\prod\limits_{s=1}^{k}
h_{\alpha_{s}^{\vee}} \prod\limits_{\delta\in\,\Phi^{+}
}^{\rightarrow} F_{\delta}^{m_{\delta}}\,q^{h_{\rho}}\,
\prod\limits_{p=1}^{k} h_{\alpha_{p}} \prod\limits_{\gamma
\in\,\Phi^{+} }^{\rightarrow} E_{\gamma}^{m_{\gamma}}\\
\nonumber
 =\sum\limits_{k=0}^{\infty}\,
\dfrac{h^k}{k!} \sum\limits_{\alpha_{1}...\,\alpha_{k}\in
\Delta}\,\sum\limits_{m_{\beta},\,\beta\in\,\Phi^{+}}^{\infty}
\prod\limits_{\eta\in\,\Phi^{+} }\,
\dfrac{(q_{\eta}-q_{\eta}^{-1})^{m_{\eta}}}{[m_{\eta}]_{q_{\eta}}!}\,
q^{m_{\eta}(m_{\eta}-1)/2}_{\eta}\,q^{(\,\rho,\,m_{\eta}\eta\,) }
\prod\limits_{s=1}^{k} h_{\alpha^{\vee}_{s}} (h_{\alpha_{s}}
+\sum\limits_{\delta\in\,\Phi^{+}}(\,\alpha_{s},\, m_{\delta
}\delta\,) ) \,q^{h_{\rho}}\\ \label{step3} \prod\limits_{\delta
\in\, \Phi^{+} }^{\rightarrow} F_{\delta}^{m_{\delta}}
\prod\limits_{\gamma\in\,\Phi^{+}}^{\rightarrow}
E_{\gamma}^{m_{\gamma}}\\ \label{step4}
=\sum\limits_{m_{\beta},\,\beta\in\,\Phi^{+}}^{\infty}
\prod\limits_{\eta\in\, \Phi^{+} }\,
\dfrac{(q_{\eta}-q_{\eta}^{-1})^{m_{\eta}}}{[m_{\eta}]_{q_{\eta}}!}\,
q^{m_{\eta}(m_{\eta}-1)/2}_{\eta}\,q^{(\,\rho,\,m_{\eta}\eta\,) }
\,q^{\sum \limits_{\alpha\in\, \Delta } h_{\alpha^{\vee}} (
h_{\alpha}+\sum\limits_{\delta\in\,\Phi^{+}}(\,\alpha,\,m_{\delta}\delta\,)
) } q^{h_{\rho}}\prod\limits_{\delta \in\Phi^{+} }^{\rightarrow}
F_{\delta}^{m_{\delta}} \prod\limits_{\gamma\,\in\,\Phi^{+}
}^{\rightarrow} E_{\gamma}^{m_{\gamma}}\\ \label{step5}
 =q^{\sum\limits_{\alpha\,\in\, \Delta
} h_{\alpha^{\vee}} h_{\alpha} } q^{h_{\rho}}
\sum\limits_{m_{\beta}, \beta\,\in\,\Phi^{+}}^{\infty}
\prod\limits_{\eta\,\in\,\Phi^{+} }
\dfrac{(q_{\eta}-q_{\eta}^{-1})^{m_{\eta}}}{[m_{\eta}]_{q_{\eta}}!}\,
q^{m_{\eta}(m_{\eta}-1)/2}_{\eta} q^{ (h_{\eta}+(\,\rho,\,
\eta\,))m_{\eta} } \prod\limits_{\delta\,\in\,\Phi^{+}
}^{\rightarrow} F_{\delta}^{m_{\delta}}
\prod\limits_{\gamma\,\in\,\Phi^{+} }^{\rightarrow}
E_{\gamma}^{m_{\gamma}}\\ \label{step6}
=q^{\sum\limits_{\alpha\,\in \,\Delta } h_{\alpha^{\vee}}
h_{\alpha} } q^{h_{\rho}} \prod\limits_{\eta\,\in\, \Phi^{+}}
\sum\limits_{m=0}^{\infty} \dfrac{(q-q^{-1})^{m}}{[m]_{q}!}
\,q^{m(m-1)/2}\, q^{(h_{\eta}+2)\,m }
E^{m}_{\eta}\,F^{m}_{\eta} \\
\label{step7}=q^{\,\sum\limits_{\alpha\,\in\, \Delta }
h_{\alpha^{\vee}} h_{\alpha} } q^{\,h_{\rho}}
\prod\limits_{\eta\,\in \,\Phi^{+}} \left\{
q^{e_{\eta}f_{\eta}+f_{\eta}e_{\eta}-h_{\eta}}\right\}= q^{\,
\sum\limits_{\alpha\,\in\, \Delta } h_{\alpha^{\vee}} h_{\alpha} +
\sum\limits_{\alpha\,\in\, \Phi^{+}}  e_{\alpha} f_{\alpha}+
f_{\alpha}e_{\alpha}  }=q^{\Omega_{2}}
 \ee
 \newpage
Here we used the following facts and identities:
 \be\nonumber
\tr_{2} \left\{ {\hat P} A\otimes B \right\}= B\,A\ \ \ \
(\ref{step1})\rightarrow(\ref{step2})\\
 \nonumber
 q^{h_{\rho}}\,F_{\delta}^{m_{\delta}}\,q^{-h_{\rho}}=q^{(\,\rho,\,\delta\,)\,m_{\delta}}F_{\delta}^{m_{\delta}},\ \ \
 \prod\limits_{\delta
 \in\Phi^{+}}^{\rightarrow}\,F_{\delta}^{m_{\delta}}\,h_{\alpha}=\left( h_{\alpha}+\sum\limits_{\delta \in \Phi^{+}} (\,\alpha,\, \delta\,m_{\delta}
 \,)
 \right)\, \prod\limits_{\delta\,\in\,\Phi^{+}}^{\rightarrow}
 F_{\delta}^{m_{\delta}}\ \ \  (\ref{step2})\rightarrow(\ref{step3})
\\
\nonumber  \sum\limits_{\alpha\,\in \,\Delta} h_{\alpha^{\vee}}
(\,\alpha,\, \eta\,) =\sum\limits_{\alpha\,\in\, \Delta}
h_{\alpha} (\,\alpha^{\vee}, \,\eta\,)=h_{\eta} \ \ \
(\ref{step4})\rightarrow(\ref{step5})
\\
\nonumber
 \sum\limits_{m=0}^{\infty}
\dfrac{(q-q^{-1})^{m}}{[m]_{q}!} q^{m(m-1)/2}q^{(h_{\eta}+2)\,m }
E^{m}_{\eta}\,F^{m}_{\eta}=q^{e_{\eta}f_{\eta}+f_{\eta}e_{\eta}-h_{\eta}
}\ \ \ (\ref{step6})\rightarrow(\ref{step7}) \ee
 %substituting this into the last equation we get:
 %$$
%q^{\sum\limits_{\alpha\in \Delta } h_{\alpha^{\vee}} h_{\alpha} }
%q^{h_{\rho}} \prod\limits_{\eta\in \Phi^{+}} \left\{
%q^{e_{\eta}f_{\eta}+f_{\eta}e_{\eta}-h_{\eta}}\right\}= q^{
%\sum\limits_{\alpha\in \Delta } h_{\alpha^{\vee}} h_{\alpha} +
%\sum\limits_{\alpha\in \Phi^{+}}  e_{\alpha} f_{\alpha}+
%f_{\alpha}e_{\alpha}  }=q^{\Omega_{2}}
%$$
and $\Omega_{2}$ is by definition the quadratic Casimir element of
the algebra $U(g)$: \be \label{Casimir}
\Omega_{2}\equiv\sum\limits_{\alpha\,\in\, \Delta }
h_{\alpha^{\vee}} h_{\alpha} + \sum\limits_{\alpha\,\in\,
\Phi^{+}} e_{\alpha} f_{\alpha}+ f_{\alpha}e_{\alpha}
 \ee~The proof for ${\cal R}^{\,-1}$ is analogous. $\square$

Now, having the solution of fundamental equations one can
construct all the special point operators. The crossing operators
are expressed trough the universal $R$-matrix by (\ref{crosspl}),
(\ref{crossmin}) and (\ref{fir}).
 To construct the turning point
operators we need to pick up the operators ${\cal M}$ , and
$\overline {\cal M}$ satisfying (\ref{adrel}). The obvious choice
is:
$$
{\cal M}\,=\overline {\cal M}\,= q^{\,h_{\rho}/2},
$$
then the rest of the turning point operators are expressed trough
these two by (\ref{tur}). Explicit expressions for these operators
are summarized in (\ref{turnings}) and (\ref{crossings}).
\section{Knot invariants \label{Ex}}
The crossing and turning point operators introduced in the
previous sections allow to calculate the ambient and regular knot
invariants. In this section we describe in detail the procedure
for computation of these invariants and give examples of these
calculations for several different choices of the gauge group and
its representation. \vspace{7mm}
\subsection{Regular knot invariants. \label{method}}
Let $D\,:\, S^{1} \stackrel{K}{\rightarrow}
\mathbb{R}^{3}\stackrel{P}{\rightarrow}\mathbb{R}^{2}$ be a
two-dimensional projection of a knot $K$ (here $P$ is the
projector to the plane). Then the image $D(S^{1})$ is a closed
planar curve with finite number of self-crossings and turning
points about some distinguished direction in the plane. To compute
the regular isotopy invariant  $<W_{\varrho}(K)>$  we should
reduce all crossings of $D( S^{1})$ to one of the 8 canonical
crossings represented in Fig.\ref{CT} (additional turning points
can appear during this process). Then, we attach to every special
point the corresponding operator, for example as in the
Fig.\ref{con}. The indices of these operators correspond to the
incoming and outgoing lines of special points, therefore, the
closed curve $D(S^1)$ naturally defines the contraction of the
indices of the operators.  For instance such a contraction for the
two-dimensional projection of the knot $5_{2}$ represented in
Fig.\ref{con} reads: \be \label{cont} <W(5_2)>={{\cal
R}_{1}}^{b\,i}_{j\,c}\,{{\cal R}_{1}}^{j\,c}_{d\,k}\,{{\cal
R}_{2}}^{g\,k}_{m\,h}\,{{\cal R}_{3}}^{d\,n}_{a\,e}\,{{\cal
R}_{4}}^{p\,f}_{g\,n}\,{{\cal M}_{1}}^{e}_{f}\,{{\cal
M}_{2}}^{m}_{p}\,{{\cal M}_{3}}^{a}_{b}\,{{\cal M}_{4}}^{n}_{i}
\ee If a Wilson loop $ W_{\varrho}(K)$ is in a representation
$\varrho\,:\, G\rightarrow End(V)$ of a gauge group $G$, then the
operators of special points Fig.\ref{Rm8} read:
 \be
 \left\{
\begin{array}{ll}
\label{turnings}
 {\cal M}_{1}={\cal
M}_{4}=\varrho\,(\, q^{h_{\rho}/2}\,)\,, \\  \\ {\cal M}_{2}={\cal
M}_{3}=\varrho\,(\,q^{-h_{\rho}/2}\,)\,,
\end{array}
\right. \ee

 \be \label{crossings} \left\{  \begin{array}{ll}
 {\cal R}_{1}= \varrho\otimes\varrho\, \Big( \, {\cal R}  \,\Big )\,, \\ \\ {\cal R}_{5}=  \varrho\otimes\varrho\, \Big(\,{\cal R}^{-1}\,\Big),  \\
\\
{\cal R}_{2}= \varrho\otimes\varrho\,\Big(\,(1\otimes
q^{h_{\rho}/2})\,{\cal R}\, (q^{-h_{\rho}/2}\otimes 1 )\,\Big)\,,\\
\\ {\cal R}_{6}= \varrho\otimes\varrho\,\Big(\,(1\otimes
q^{h_{\rho}/2})\,{\cal R}^{-1}\, (q^{-h_{\rho}/2}\otimes 1
)\,\Big)\,,\\
\\
{\cal R}_{3}= \varrho\otimes\varrho\,\Big(\,(q^{h_{\rho}/2}\otimes
1)\,{\cal R}\, (1 \otimes q^{-h_{\rho}/2} )\,\Big)\,, \\ \\
{\cal R}_{7}= \varrho\otimes\varrho
\,\Big(\,(q^{h_{\rho}/2}\otimes
1)\,{\cal R}^{-1}\, (1 \otimes q^{-h_{\rho}/2} )\,\Big)\,,  \\
\\
 {\cal
R}_{4}= \varrho\otimes\varrho \,\Big((q^{h_{\rho}/2}\otimes
q^{h_{\rho}/2})\,{\cal R}\, (q^{-h_{\rho}/2} \otimes
q^{-h_{\rho/2}} )\,\Big)\,, \\ \\ {\cal R}_{8}=
\varrho\otimes\varrho\,\Big(\,(q^{h_{\rho}/2}\otimes
q^{h_{\rho}/2})\,{\cal R}^{-1}\, (q^{\,-h_{\rho}/2} \otimes
q^{-h_{\rho}/2})\, \Big)\,,
\end{array} \right.
 \ee  where ${\cal
R}$ is the universal quantum $R$-matrix  (\ref{quR}) for
$U_{h}(g)$ .
 For example in the case of $G=SU(2)$ and $\varrho$ is the
fundamental representation, the contraction (\ref{cont}) gives:
$$
<W(5_{2})>=q^{-17}\,(q+q^{-1})\,(\,
{q}^{10}-{q}^{8}+2\,{q}^{6}-{q}^{4}+{q}^{2}-1\,)
$$
\vspace{7mm}
 \bfig \bc
\includegraphics[scale=0.3]{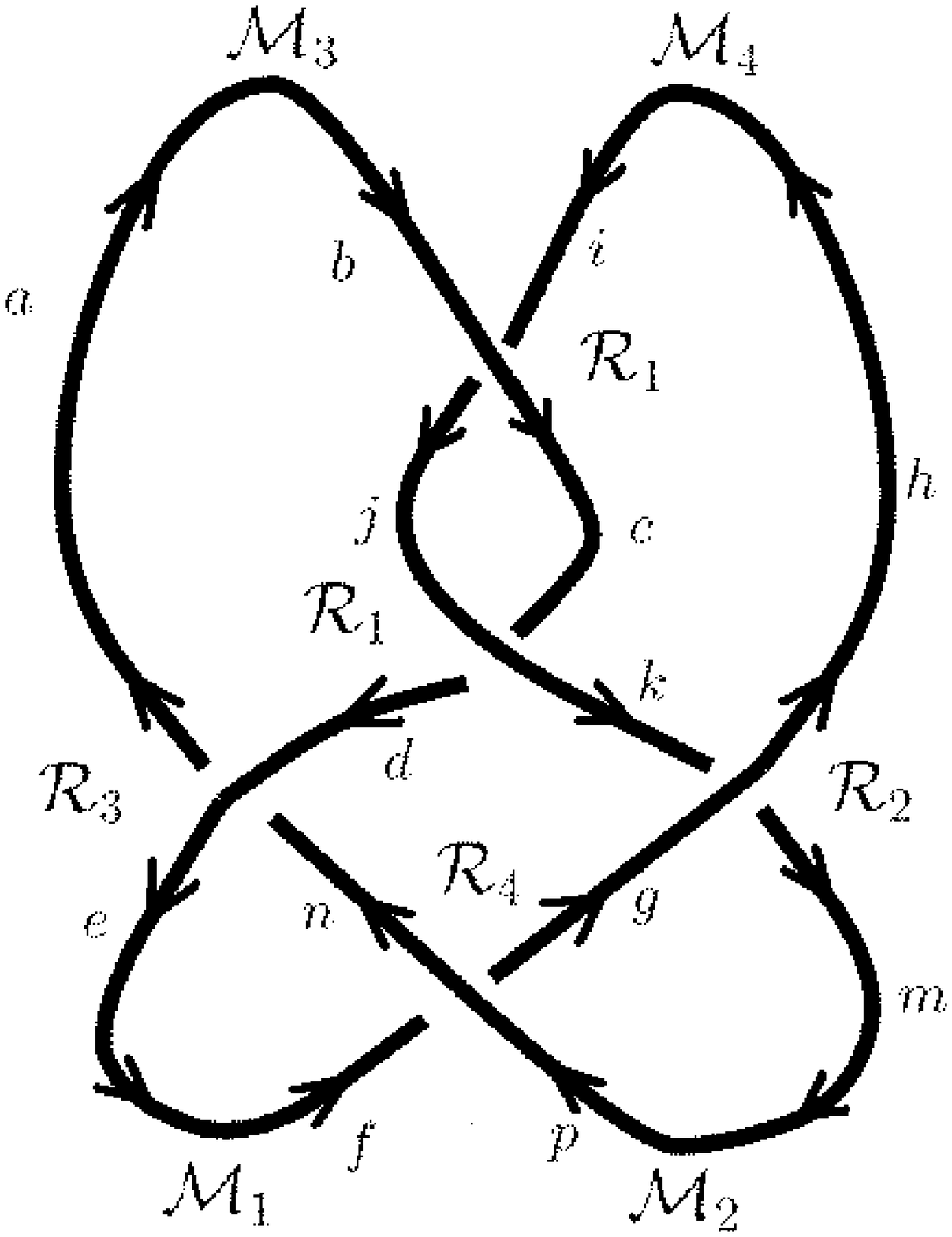}  \caption{\label{con}} \ec
\efig
 \vspace{5mm}
\subsection{Writhe number. Ambient invariants. \label{wn}}
As it was discussed in the section \ref{framing} the quantity
$<W_{\varrho}(K)>$ can be considered as an invariant of embedding
${\cal K} \,:\, S^{1}\times [\,0,1\,]\rightarrow \mathbb{R}^{3}$
of a ribbon $ [0,1]\times S^{1}$ into three-dimensional space. The
knot $K$ is defined as $K={\cal K}(x,0)\,:\,S^{1}\rightarrow
\mathbb{R}^{3}$  and  $K^{'}={\cal K}(x,1)\,:\,S^{1}\rightarrow
\mathbb{R}^{3}$ is a framing contour of $K$ introduced by
regularization of CST. The quantity $<W_{\varrho}(K)>$ depends on
the number of twists on a ribbon Fig.\ref{kn}(a), this results in
the fact that it is not an ambient isotopy invariant of the knot,
but only a regular invariant. Nevertheless, we can easily
construct the ambient isotopy knot invariant using the quantity
$<W_{\varrho}(K)>$. Indeed, let $w ({\cal K})$ is the total number
of twists of a framing ribbon embedded in $\mathbb{R}^{3}$, this
quantity is referred to as a \textit{writhe} of the ribbon.
According to (\ref{twist}), additional twist on the ribbon changes
the vev of a Wilson loop $<W_{\varrho}(K)>$ for a factor
$q^{\varrho( \Omega_{2} )}$. Therefore, the quantity
$I(G,\,\varrho,\,K)$ defined as: \be \label{ambinv}
I(G,\,\varrho,\, K)=\dfrac{1}{{<W(U_{0})>}}\,q^{-w ({\cal
K})\,\varrho( \Omega_{2} )}\,<W_{\varrho}(K)> \ee is an ambient
isotopy invariant. Here
$<W(U_{0})>$ stands for the vev of unknot and we divide by its value for normalization. \\
\indent In the temporal gauge $w ({\cal K})$ is the total number
of twists of a ribbon in the temporal framing introduced in the
section \ref{framing}. The temporal framing is the procedure that
naturally defines the class of ribbon embeddings $\{{\cal K}\}$
for a given two-dimensional projection $D\,:\, S^{1}
\stackrel{K}{\rightarrow}
\mathbb{R}^{3}\stackrel{P}{\rightarrow}\mathbb{R}^{2}$ of the knot
$K$. Therefore, one can define the writhe of a knot projection as
the writhe of its temporal framing ribbon: \be \label{wr} w(D)=w(
\cal K) \ee It is rather obvious, that this definition is correct,
i.e. it does not depend on the choice of representative framing
ribbon. As a function of $D$ the writhe (\ref{wr}) has a
constructive representation in terms of $D$ itself. Indeed, every
crossing of $D$ is equivalent up to the planar rotations to one of
the two crossings represented in  Fig.\ref{writhe}\\
 \vspace{5mm}
 \bfig \bc
\includegraphics[scale=0.5]{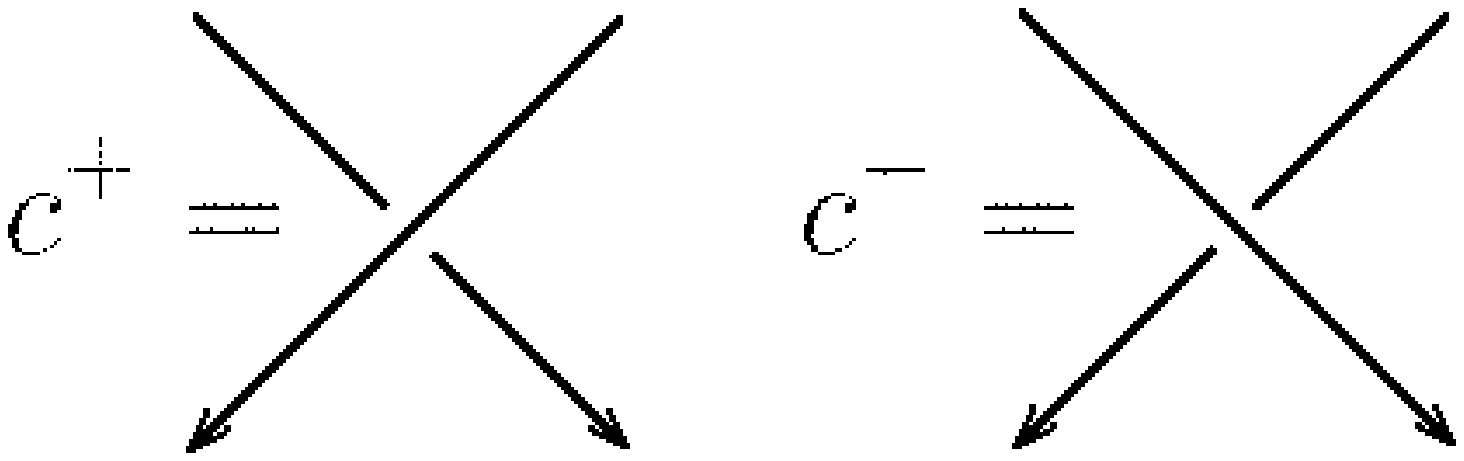}  \caption{\label{writhe}} \ec
\efig
 \vspace{5mm}
 Define the writhes of these crossings as \be w(c^{+})=+1,\ \
\ w(c^{-})=-1 \ee Then the writhe (\ref{wr}) can be expressed as
sum of writhes: \be w(D)=\sum\limits_{i} w (i) \ee where sum runs
over all crossing of $D$. \vspace{7mm}
\subsection{Braid representation of the knots.}
A useful way for representing knots is to make use the braid group
$B_{n}$. The Artin braid group $B_{n}$ (the braid group
corresponding to the $A_{n}$ root system in terms of section
\ref{sub3}) of $n$ strings is the group generated by $n-1$
generators $g_{1}, g_{2},...,g_{n-1}$ satisfying the following
relations: \be \nonumber g_{i}\,g_{j}=g_{j}\,g_{i} \ \ \ \
\textrm{for} \ \ \ |i-j|>1 \ \ \ \ \ \ \
\\ \label{b1}
\\ \nonumber
 g_{i}\,g_{i+1}\,g_{i}=g_{i+1}\,g_{i}\,g_{i+1}, \ \ \ \ \textrm{for} \ \ \ \ i-1<n \ee
Let $\varrho \,:\, U_{h}(g) \rightarrow End(V)$ be a
representation of some quantum universal enveloping algebra, then
one can construct associated representation of  the braid group
$\varrho_{B_{n}}\,:\,B_{n}\rightarrow End( V^{\otimes n} )$ by the
following explicit definition:
$$
\varrho_{B_{n}} \,:\, g_{i} \mapsto \underbrace{1\otimes1
\otimes...\otimes1}_{1...i-1}\,\otimes\,  \textsf{R}_{\varrho}
\otimes \underbrace{ 1 \otimes...\otimes1}_{i+2...n},\ \ \ \
\textsf{R}_{\varrho}= \varrho\otimes\varrho\,({\cal R})
$$
The first defining relation (\ref{b1}) in this representation is
obvious, and the second one simply follows from~QYBE~(\ref{QYBE}).\\
\indent
 Each elements of $B_{n}$ admits a graphical representation. One can
 associate to the element $g_{i}\in B_{n}$ the picture shown in
 Fig.\ref{br} (a), then the element $g_{1}\,g_{2}\,g_{1}^{-1} \in
 B_{n}$ can be represented as in the Fig.\ref{br}(b):
\bfig \bc
\includegraphics[scale=0.8]{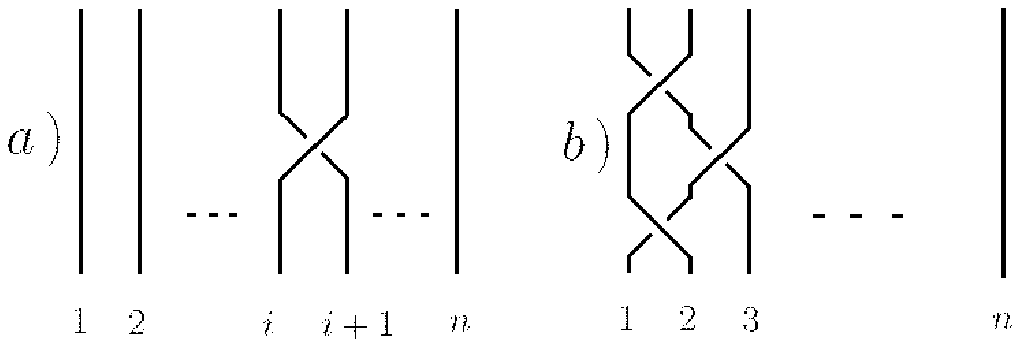}  \caption{\label{br}} \ec
\efig \noindent
 i.e. $g_{i}$ can be considered as a twist of $i$ and $i+1$
 string on the set of $n$ strings. The closure ${\hat b}$ of the
 element $b\in B_{n}$ is a two-dimensional diagram obtained by
 connection the endpoints of the strings of $B_{n}$ in a order
 preserving way Fig.\ref{cl} (a):

\bfig \bc
\includegraphics[scale=0.8]{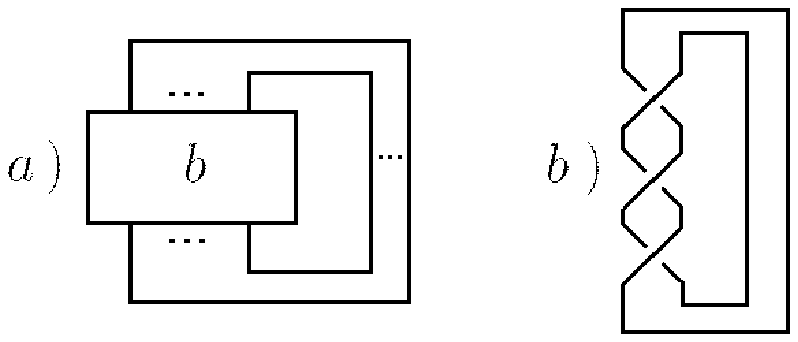}  \caption{\label{cl}} \ec
\efig

 Every link and knot in particular, can be represented as a
 closure of an element $b\in B_{n}$
 (this element and group $B_{n}$ are not uniquely defined). For
 instance the simplest knot $3_{1}$ can be represented as a
 closure of $ g_{1}^{3} \in B_{2}$ Fig.\ref{cl}(b). The braid
 representation of knots provides useful tool for calculation of knot invariants
 (\ref{ambinv}).
 Indeed, applying the method for calculation of $I(G,\,\varrho,\,K)$ described in the
 sections \ref{method}-\ref{wn}
 to a closure of $b \in B_{n}$ we arrive to the following
 proposition:
\\
 \indent
 \textit{Let knot $K$ be represented as a
closure $\hat b$ of some element of a braid group $b\in B_{n}$.
Then an ambient isotopy invariant of a knot for gauge group $G$
and representation $\varrho: g\rightarrow \textrm{End}(V) $ is
given by the following explicit formula: \be \label{brep}
\begin{array}{|c|} \hline
\\
 I_{\varrho}(G, \,
\varrho,\,K)=\dfrac{1}{\tr(\textsf{Q}_{\varrho} )}\,q^{-w({\hat
b})\, \varrho(\Omega_{2}) }\,\tr( \textsf{Q}^{\otimes n}_{\varrho}
\, \textsf{b}_{\varrho} )\\  \\ \hline \end{array} \ee where
$w({\hat b})$ is a writhe of the diagram associated with ${\hat
b}$ and \be \textsf{Q}_{\varrho}=\varrho\, ( {\cal Q} ),\ \ \
\textsf{b}_{\varrho}=\varrho_{B_{n}}( b ),\ \ \ \tr(
\textsf{Q}_{\varrho} ) =<W(U_{0})>_{\varrho} \ee
 }
In the braid representation of a knot ${\hat b}$, every string of
$b$ has only two turning points with respect to the vertical
direction of the page. This is why the contribution of turning
points appears in (\ref{brep}) in the form ${\cal Q}= {\cal M}
\overline{\ {\cal M} }= {\cal M}^2$. The calculation of
(\ref{brep}) for some particular knot is just a matter of
multiplication and taking a trace of relatively big matrices. In
the next subsection we give several explicit examples of such
calculations of the knot invariants for first five nontrivial
knots. We also give explicit expressions for ${\cal R}$ and ${\cal
Q}$ for different groups and their representations.
\subsection{Explicit examples}
In this section we give examples of explicit calculations of knot
invariants and universal $R$-matrix (\ref{quR}) for several
particular groups and their representations. The invariants are
calculated for the first five non-trivial knots from Rolfsen table
\cite{katlas} : $3_1,\,4_1,\,5_{1}, 5_{2}$ and $6_{1}$. The two
dimensional projections of these knots with the temporal framing
contour can be chosen as in the Fig.\ref{knots}:

\bfig \bc
\includegraphics[scale=0.3]{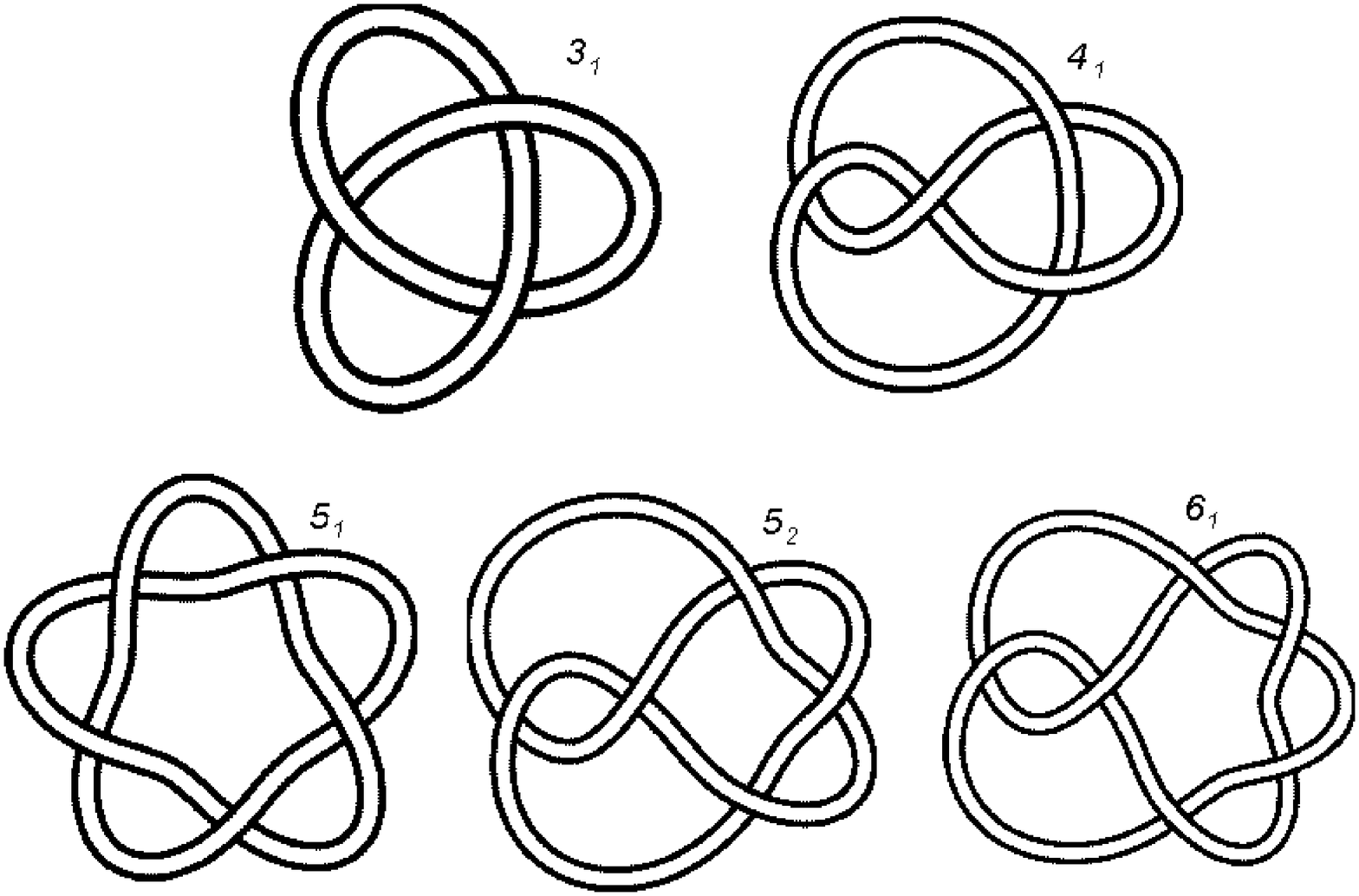}  \caption{\label{knots}} \ec
\efig \noindent
 The braid representations of these knots and the writhe
numbers of the corresponding closures are summarized in the table:
\be \label{kntable}
\begin{array}{|l|l|l|}\hline \textrm{Knot}& \textrm{Braid representation} &  \textrm{Writhe} \\ \hline 3_1 & b=g_{1}^{3}\in B_{2} & w({\hat b})=3 \\
\hline 4_{1} & b=g_{2}^2\, g_{1}^{-1}\, g_{2} \, g_{1}^{-1}\in
B_{3} & w( {\hat b} )=1\\ \hline 5_{1} & b=g_{1}^{5}\in B_{2} &
w({\hat b})=5\\ \hline 5_{2} &
b=g_{2}^{3}\,g_{1}\,g_{2}^{-1}\,g_{1} \in B_{3} & w(\hat b)=4\\
\hline 6_{1} & b=
g_{1}\,g_{2}^{-1}\,g_{3}\,g_{1}\,g_{2}^{-1}\,g_{3}^{-2}\in B_{4} &
w({\hat b})=-1\\ \hline
\end{array} \ee\\

\vspace{5mm}
\subsubsection{$SU(2)$ representations of the weight $\lambda$.}
The polynomial knot invariants arising from vevs of Wilson loops
carrying arbitrary representation of the group $SU(2)$ is a very
instructive example for application of general formulas of
sections \ref{top}-\ref{Sol}. For this reason let us consider in
detail the construction of quantum generators for $U_{h}(su(2))$
and universal $R$-matrix (\ref{quR}) in the case of representation
with highest weight $\lambda$.\\
\indent The algebra $su(2)$ is the tree-dimensional Lie algebra
with generators $e$, $f$ and $h$ subjected to the following
relations:
$$
[h,e]=2e,\ \ \ [h,f]=-2f,\ \ \ [e,f]=h
$$
All irreducible finite-dimensional representations of $su(2)$ is
the representations of highest weight $\varrho_{\lambda}:
su(2)\rightarrow End(V_{\lambda}),\ \ \lambda\in \mathbb{N}_{0}$.
The dimension of the representation with highest weight $\lambda$
is given by $\dim(V_{\lambda})=\lambda+1$. In a suitable basis in
$V_{\lambda}$ the generators $e$, $f$ and $h$ are represented by
$(\lambda+1)\times(\lambda+1)$ matrices with the following
elements: \be
\begin{array}{ll}
 ( \varrho_{\lambda}\,e )_{i j}:=(e_{\lambda})_{i
j}=( \lambda-i+1 ) \,\delta_{i, j-1} & \\ &  \\ (
\varrho_{\lambda}\,f
)_{i j}:= (f_{\lambda})_{i j}=i \, \delta_{i-1, j}\, &  i,j=1...\lambda+1\\ &  \\
 ( \varrho_{\lambda}\,h )_{i j}:=(h_{\lambda})_{i j}=\Big(
\lambda-2(i-1) \Big)\,\delta_{i,j}\, & \end{array}
 \ee
To get from Lie algebra generators of $su(2)$ to the quantum group
generators $E$, $F$ and $h$ we replace the matrix elements of $e$
and $f$ by their quantum deformations: \be
\begin{array}{|ll|}
\hline &
 \\ (E_{\lambda})_{i
j}=[ \,\lambda-i+1\, ] \,\delta_{i, j-1}, & \\ &  \\ (F_{\lambda})_{i j}=[\,i\,] \, \delta_{i-1, j}\, &  i,j=1...\lambda+1\\ &  \\
(h_{\lambda})_{i j}=\Big( \lambda-2(i-1) \Big)\,\delta_{i,j}\,. & \\ & \\
\hline
\end{array}
 \ee
 where
 $$
  \\
{[}m{]}=\dfrac{q^{m}-q^{-m}}{q-q^{-1}}
 $$
 It is a simple exercise to check that for arbitrary $\lambda$
 these matrices satisfy the relations (\ref{qserr1})-(\ref{qserr4}) for $U_{h}(su(2))$:
 \be
 [\,h_{\lambda}, E_{\lambda}\,]=2\,E_{\lambda}\,, \ \ \ [\,h_{\lambda},F_{\lambda}\,]=-2\,F_{\lambda}\,,\ \ \
 [\,E_{\lambda},F_{\lambda}\,]=\dfrac{q^{h_{\lambda}}-q^{-h_{\lambda}}}{q-q^{-1}}\,
 \ee
 Note, that in the representation of weight $\lambda$ the
 generators $E_{\lambda}$ and $F_{\lambda}$ are subjected to
 the  additional relation:
 \be
 E_{\lambda}^{\lambda+1}=F_{\lambda}^{\lambda+1}=0
 \ee
 Therefore, in this case, the universal quantum $R$-matrix (\ref{quR})
 is given by the following finite sum:
 \be  \textsf{R}_{\lambda}=\varrho_{\lambda}\otimes\varrho_{\lambda} \,({\cal R}) ={\hat P}\, q^{\,h_{\lambda}\otimes h_{\lambda}/2}
 \sum\limits_{m=0}^{\lambda+1}\,\dfrac{(q-q^{-1})^{m}}{[\,m\,]!}\,q^{m(m-1)/2}
 F_{\lambda}^{m} \otimes E_{\lambda}^{m} \ee
 The operator (\ref{Q}) in this case reads:
 \be  \textsf{Q}_{\lambda}= \varrho_{\lambda} ({\cal Q})= q^{h_{\lambda}}= \textrm{diag}[\,q^{\lambda-2(i-1)}\,]\,,\ \ i=1...\lambda+1 \ee
 For the quadratic Casimir element (\ref{Casimir}) we find:
 \be  \varrho_{\lambda}( \Omega_{2}) = \frac{1}{2}\,h_{\lambda}\,h_{\lambda}+ e_{\lambda}\,f_{\lambda}+f_{\lambda}\,e_{\lambda} = \lambda(\lambda+2) \ee
Note that for vev of the unknot we have: \be  <W(U_{0})> =
\textrm{tr} \,\textsf{Q}_{\lambda}= \textrm{tr}\, q^h_{\lambda}
=\sum\limits_{m=1}^{\lambda+1} q^{\lambda-2(m-1)}   \ee Therefore,
the ambient isotopy invariant (\ref{brep}) for a knot, represented
by the closure of $b\in B_{n}$, is: \be
\begin{array}{|l|}
\hline \\
 I( SU(2), \lambda, K
) =\, \dfrac{1}{ \sum\limits_{m=1}^{\lambda+1}\,
q^{\,\lambda-2(m-1)} }\, q^{ -w(K)\,\lambda(\lambda+2) }
\,\textrm{tr}\,\left( \textsf{Q}_{\lambda}^{ \otimes n} \,
\textsf{b}_{\lambda}(K) \right)  \\ \\  \hline \end{array}\ee
Using the braid representations for first six non-trivial knots
and their writhe numbers in the standard framing (\ref{kntable})
we find that for fundamental representation $\lambda=1$ this
invariant gives the Jones polynomials of the knots: \be
\begin{array}{ll}I(SU(2),\,1,\,3_{1})\,: &   ({q}^{6}+{q}^{2}-1)\,{q}^{-8} \\
\\
 I(SU(2),\,1,\,4_{1})\, : & ({q}^{8}-{q}^{6}+{q}^{4}-{q}^{2}+1){q}^{-4} \\
 \\
I(SU(2),\,1,\,5_{1})\, : &
({q}^{10}+{q}^{6}-{q}^{4}+{q}^{2}-1){q}^{-14}\\
\\
I(SU(2),\,1,\,5_{2})\, : &
({q}^{10}-{q}^{8}+2\,{q}^{6}-{q}^{4}+{q}^{2}-1){q}^{-12}\\
\\
I(SU(2),\,1,\,6_{1})\, : &
({q}^{12}-{q}^{10}+{q}^{8}-2\,{q}^{6}+2\,{q}^{4}-{q}^{2}+1){q}
^{-4}
 \end{array}\ee
For a general value of the highest weight $\lambda$ we find: \be
\label{SU2poly}
\begin{array}{ll} I(SU(2),\,\lambda,\,3_{1})\,: &  \dfrac{1}{[\lambda+1]} \sum\limits_{m=0}^{\lambda}
 [\,2m+1\,]\,(-)^{\lambda-\,m}\,q^{ \,3(\, \lambda( \lambda+2) -m(m+1)\, ) } \\
\\
 I(SU(2),\,\lambda,\,4_{1})\, : &  \dfrac{1}{[\lambda+1]} \sum\limits_{m,\, k=0}^{\lambda} \,\sqrt{ [\,2m+1\,][\, 2k+1 \,]  }\, a_{k\,m} \,q^{ \,2(\, m(m+1)-k(k+1)\, ) } \\
 \\
I(SU(2),\,\lambda,\,5_{1})\, : &
 \dfrac{1}{[\lambda+1]} \sum\limits_{m=0}^{\lambda}
 [\,2m+1\,]\,(-)^{\lambda-\,m}\,q^{ \,5(\, \lambda( \lambda+2) -m(m+1)\, ) }\\
\\
I(SU(2),\,\lambda,\,5_{2})\, : &
  \dfrac{1}{[\lambda+1]} \sum\limits_{m,\, k=0}^{\lambda} \,\sqrt{ [\,2m+1\,][\, 2k+1 \,]  }\, a_{k\,m} \,q^{ \,2(\,\lambda(\lambda+1)-m(m+1)-3k(k+1)/2\, ) }   \\
\\
I(SU(2),\,\lambda,\,6_{1})\, : & \dfrac{1}{[\lambda+1]}
\sum\limits_{m,\, k=0}^{\lambda} \,\sqrt{ [\,2m+1\,][\, 2k+1 \,]
}\, a_{k\,m} \,q^{ \,2(\, m(m+1)-2k(k+1)\, ) }
 \end{array}\ee
where the following conventions are used: \be
[\,m\,]=\dfrac{q^m-q^{-m}}{q-q^{-1}}\,,\
 \ \  a_{k\,m}=
(-)^{k+m-\lambda}\,\sqrt{ [\,2k+1\,][\,2m+1\,] }\, \left(
\lambda/2 ,\, \lambda/2, \,k \atop \lambda/2,\,\lambda/2,\, m
\right) \ee  The quantum Recah coefficients are: \be \nonumber
\left( p_{1} ,\, p_{2}, \,p_{12} \atop p_{3},\,p_{4},\, p_{23}
\right) = \Delta
(p_{1},\,p_{2},\,p_{12})\,\Delta(p_{3},\,p_{4},\,p_{12})\,\Delta(p_{1},\,p_{4},\,p_{23})\,\Delta(p_{3},\,p_{2},\,p_{23})\times
 \\ \nonumber   \\ \nonumber \times\sum\limits_{m>0}  (-)^{m}[m+1]! \,\, \left\{[\, m-{ p_1}-{ p_2}-{ p_{12}}]!
  [\,m-{p_{3}}-{ p_{4}}-{p_{12}}\,]!\,
 {[\,m-{p_{1}}-{ p_{4}}-{ p_{23}}
 \,]!} \times\, \right. \\ \\ \nonumber { \times [\,m-{ p_{3}}-{ p_{2}}-{ p_{23}}\,]!}\,
 {[\,{ p_1}+{ p_2}+{ p_3}+{ p_4}-m\, ]!}\,  [\,{
p_1}+{ p_3}+{ p_{12}}+{ p_{23}}-m\,]! \times\,\\
\nonumber
\\ \nonumber \left.{ \times[\,{ p_{2}}+{ p_{4}}+{ p_{12}}+{
p_{23}}-m\,]!}\right\}^{-1}
   \ee
where $m$ runs over all non-negative numbers, such that each
q-factorial in the sum gets non-negative argument, and: \be
   \Delta(a,\,b,\,c )=\sqrt{\dfrac{ [-a+b+c]![a-b+c]![a+b-c]!  }{[a+b+c+1]!}}
\ee The polynomials (\ref{SU2poly}) were obtained in
\cite{su2polynoms}  as traces of monodromies for the correlators
of the associated $SU(2)$ Wess-Zumino conformal field theory.
\subsubsection{$SU(N)$ in fundamental representation}
The irreducible representations of $su(N)$ are representations of
the highest weight $\lambda=\{
\lambda_{1}\geq\lambda_{2}\geq...\geq\lambda_{N-1} \},\ \
\lambda_{i}\in\mathbb{N}_{0}$. Here we consider the case of the
first fundamental representation corresponding to
$\lambda_{f}=\{1,0...0\}$. In this fundamental representation the
root elements of the quantum group $U_{h}(su(N))$ have the
following property:
$$
\varrho_{\lambda_{f}}(E_{\alpha})^{2}=\varrho_{\lambda_{f}}(F_{\alpha})^{2}=0,\
\ \ \alpha\in \Phi^{+}
$$
and the universal $R$-matrix (\ref{quR}) takes the form:
$$
\textsf{R}_{\lambda_{f}}= {\hat P}\,
\varrho_{\lambda_{f}}\otimes\varrho_{\lambda_{f}}( {\cal R} )=
{\hat P}\, q^{\, \sum\limits_{ \alpha\in\, \Delta }
\varrho_{\lambda_{f}}\otimes\varrho_{\lambda_{f}}(
h_{\alpha}\otimes h_{\alpha^{\vee}})} \prod\limits_{\alpha\in
\,\Phi^{+}} \Big( 1+(q-q^{-1})
\varrho_{\lambda_{f}}\otimes\varrho_{\lambda_{f}}(F_{\alpha}\otimes\,E_{\alpha})
\Big)
$$
In the standard basis $e_{ij}$ of $End(\mathbb{R}^{N})$ this
matrix can be expressed as:
$$
\textsf{R}_{\lambda_{f}}= {\hat P}\,\left(q\,\sum\limits_{i=1}^{N}
e_{i\,i} \otimes e_{i\,i}\,+ \sum\limits_{i \neq j}
e_{i\,i}\otimes e_{j\,j}\,+\sum\limits_{i>j} (q-q^{-1})\,
e_{i\,j}\otimes e_{j\,i})\right), \ \ \ \textrm{where} \ \ \ {\hat
P}=\sum\limits_{i,j=1}^{N} \,e_{ij}\otimes e_{ji}
$$
The element (\ref{Q}) in the fundamental representation reads:
$$
\textsf{Q}_{\lambda_{f}}=\varrho_{\lambda_{f}}({\cal Q})=
\textrm{diag}( q^{-N+2\,i-1} ),\ \ \ i=1..N
$$
The quadratic Casimir element (\ref{Casimir}) is :
$$
\varrho_{\lambda_{f}}(\Omega_{2})= \sum\limits_{i,j=1}^{N}
\,e_{i\,j}\,e_{j\,i} = N
$$
For the vev of the unknot in this representation we find: \be
<W(U_{0})>\,=\tr\,\textsf{Q}_{\lambda_{f}}=\sum\limits_{m=1}^{N}\,
q^{-N+2\,m-1}=\dfrac{q^{N}-q^{-N}}{q-q^{-1}} \ee Therefore the
ambient isotopy invariant
(\ref{brep}) in this case reads: \be \begin{array}{|c|} \hline \\
I(SU(N),\, \lambda_{f},\, K )= \dfrac{q-q^{-1}}{q^{N}-q^{-N}}
\,q^{-N\,w(K)}\,\tr( \textsf{Q}_{\lambda_{f}}^{\otimes n}\,
\textsf{b}_{\lambda_{f}} (K) ) \\ \\ \hline \end{array} \ee
 Using the braid representations (\ref{kntable}) for first six non-trivial knots
 and their writhe numbers in the temporal framing we get: \be
\begin{array}{ll}
I(SU(N),\, \lambda_{f},\,3_{1})\, :& \theta \left(
1+{q}^{4}-\theta\,{q}^{4} \right)\\
\\
I(SU(N),\, \lambda_{f},\,4_{1})\, :& (1-{q}^{-2}+{\theta ^{-1}
{q}^{2}} -
{q}^{2}+\theta\,{q}^{2})\\
\\
I(SU(N),\, \lambda_{f},\,5_{1})\, :&\,{\theta}^{2} \left(
1+{q}^{4}-\theta\,{q}^{4}+{q}^{8}-\theta\,{q}^{8} \right)\\
\\
I(SU(N),\, \lambda_{f},\,5_{2})\, :& \, \theta\, \left(
1-{q}^{2}+\theta\,{q}^{2}+{q}^{4}-\theta\,{q}^{4}+\theta\,{q}^{6}-{\theta}^{2}{q}^{6}
 \right)\\
 \\
 I(SU(N),\, \lambda_{f},\,6_{1})\,: &\, {\theta}^{-2}{q}^{-4} ({1-\theta+\theta{q}^{2}-{\theta}^{2}{q}^{2}-\theta{q}^{4}+2\,{\theta}^{2}{q}^{4}
 -{\theta}^{2}{q}^{6}+{\theta}^{3}{q}^{6}})
\end{array}\ee
where $\theta=q^{-2 N-2}$. In accordance with \cite{WCS} these
invariants are the HOMFLY polynomials \cite{HOMFLY} of the knots.
\subsubsection{$SO(N)$ in fundamental representation}
The root system of so(N) corresponds to $B_{N/2}$  for even $N$
and to $D_{(N-1)/2}$ for even N. In both cases the root system
consist of a roots of two types with lengthes $(\alpha,\alpha)=2$
(long roots) and $(\alpha,\alpha)=1$ (short roots).
 In the fundamental representation  the root elements of the
 algebra $U_{h}(so(N))$ are nilpotent:
$$
\varrho_{\lambda_{f}} (E_{\alpha})^{2}=\varrho_{\lambda_{f}}
(F_{\alpha})^{2}=0,\ \textrm{if $\alpha$ is long}, \ \ \
\varrho_{\lambda_{f}} (E_{\alpha})^{3}=\varrho_{\lambda_{f}}
(F_{\alpha})^{3}=0, \ \textrm{if $\alpha$ is short},
$$
The universal $R$-matrix (\ref{quR}) takes the form:
$$
\textsf{R}_{\lambda_{f}}= {\hat P}\, \varrho_{\lambda_{f}} \otimes
\varrho_{\lambda_{f}}({\cal R}) = q^{\, \sum\limits_{ \alpha\in\,
\Delta } \varrho_{\lambda_{f}} \otimes \varrho_{\lambda_{f}}
(h_{\alpha}\otimes h_{\alpha^{\vee}})} \prod\limits_{\alpha\in
\,\Phi^{+}}^{\rightarrow} \sum
\limits_{m=0}^{2}\dfrac{q^{m(m-1)/2}}{[m]_{q_{\alpha}}!}\left(
(q_{\alpha}-q_{\alpha}^{-1})^m\,
\varrho_{\lambda_{f}}(F_{\alpha})^m\otimes\,\varrho_{\lambda_{f}}(E_{\alpha})^m
 \right)
$$
The explicit expression for quantum R-matrix is different for even
and odd $N$. In suitable basis for $N=2\,n+1$ we have: \be
\textsf{R}_{\lambda_{f}}={\hat P}\, (e_{(N+1)/2\,(N+1)/2}\otimes
e_{(N+1)/2\,(N+1)/2} +\sum\limits_{i\neq N+1-i}^{N} (
\,q\,e_{i\,i}\otimes e_{i\,i}+q^{-1}e_{i\,i}\otimes
e_{N+1-i\,N+1-i}\,)+ \atop \sum\limits_{i\neq j \atop i\neq
N+1-j}^{N} e_{i\,i}\otimes e_{j\,j}+\sum \limits_{i>j}^{N}\left(
(q-q^{-1})e_{i\,j}\otimes
e_{j\,i}+(q-q^{-1})\,q^{\nu_{i}-\nu_{j}}\,e_{i\,j}\otimes
e_{N+1-i\,N+1-j}\,\right))\ee
 where
$$
\nu_{i}=\left\{
\begin{array}{c}
n+1/2-i,  i<n+1 \\0,\ \ \ \ \ \ \ \ \ \ \ \ \ \ i=n+1 \\
n+3/2-i, i>n+1
\end{array}\right.
$$
and

$$
{\hat P}=  \sum\limits_{i,j=1}^{N} \,e_{ij}\otimes e_{ji}
$$
 In the case of even $N=2n$ we get: \be  \textsf{R}_{\lambda_{f}}={\hat P}\,\sum\limits_{i\neq
N+1-i}^{N} ( \,q\,e_{i\,i}\otimes e_{i\,i}+q^{-1}e_{i\,i}\otimes
e_{N+1-i\,N+1-i}\,)+ \atop +\sum\limits_{i\neq j \atop i\neq
N+1-j}^{N} e_{i\,i}\otimes e_{j\,j}+\sum \limits_{i>j}^{N}\left(
(q-q^{-1})e_{i\,j}\otimes
e_{j\,i}+(q-q^{-1})\,q^{\nu_{i}-\nu_{j}}\,e_{i\,j}\otimes
e_{N+1-i\,N+1-j}\,\right) \ee where
$$
\nu_{i}=\left\{
\begin{array}{c}
n-i,  \ \ \ \ i<n+1\\
n-i+1, i\geq n+1
\end{array}\right.
$$
The operator (\ref{Q}) in both cases is given by the following
expression:
$$
\textsf{Q}_{\lambda_{f}}=\varrho_{\lambda_{f}}({\cal
Q})=\textrm{diag}(q^{\,2\,\nu_{\,N+1-i}})
$$
The quadratic Casimir element:
$$
\varrho_{\lambda_{f}}(\Omega_{2})=\sum\limits_{\alpha\in \Delta }
h_{\alpha^{\vee}} h_{\alpha} + \sum\limits_{\alpha\in \Phi^{+}}
e_{\alpha} f_{\alpha}+ f_{\alpha}e_{\alpha}=N-1
$$
For the vev of unknot we get: \be
<W(U_{0})>=\,\sum\limits_{m=1}^{N}\, q^{\,2 \nu_{\,N+1-i}} \ee
Therefore the ambient isotopy invariant (\ref{brep}) in this case
reads: \be \begin{array}{|c|} \hline \\  I(SO(N),\, \lambda_{f},\,
K )= \dfrac{1}{\sum\limits_{m=1}^{N}\, q^{\,2\nu_{\,N+1-i}}}
\,q^{-(N-1)\,w(K)}\,\tr( \textsf{Q}_{\lambda_{f}}^{\otimes n}\,
\textsf{b}_{\lambda_{f}} (K) )\\ \\   \hline \end{array} \ee
 Using the braid representations (\ref{kntable}) for first six non-trivial knots
 and their writhe numbers in the temporal framing we get:
 \be
 \label{Kauff}
 \begin{array}{ll}
 I(3_{1})\,:& 2\,{a}^{2}-{a}^{4}+ \left( -{a}^{3}+{a}^{5} \right) z+ \left( {a}^{2}-
{a}^{4} \right) {z}^{2}\\
\\
I(4_{1})\,:&-1+{a}^{-2}+{a}^{2}+ \left( {a}^{-1}-a \right) z+
\left( -2+{a}^{-2}+{ a}^{2} \right) {z}^{2}+ \left( {a}^{-1}-a
\right) {z}^{3}\\
\\
I(5_1)\,:&3\,{a}^{4}-2\,{a}^{6}+ \left(
-2\,{a}^{5}+{a}^{7}+{a}^{9} \right) z+
 \left( 4\,{a}^{4}-3\,{a}^{6}-{a}^{8} \right) {z}^{2}+ \left( -{a}^{5}
+{a}^{7} \right) {z}^{3}+ \left( {a}^{4}-{a}^{6} \right) {z}^{4}\\
\\
I(5_2)\,:&{a}^{2}+{a}^{4}-{a}^{6}+ \left( -2\,{a}^{5}+2\,{a}^{7}
\right) z+
 \left( {a}^{2}+{a}^{4}-2\,{a}^{6} \right) {z}^{2}+ \left( {a}^{3}-2\,
{a}^{5}+{a}^{7} \right) {z}^{3}+ \left( {a}^{4}-{a}^{6} \right)
{z}^{4 }\\
\\
I(6_{1})\,:&{a}^{-4}-{a}^{-2}+{a}^{2}+ \left(
2\,{a}^{-3}-2\,{a}^{-1} \right) z+
 \left( 3\,{a}^{-4}-4\,{a}^{-2}+{a}^{2} \right) {z}^{2}+ \left( 3\,{a}
^{-3}-2\,{a}^{-1}-a \right) {z}^{3} +\\  \\  & + \left(
1+{a}^{-4}-2\,{a}^{-2}
 \right) {z}^{4}+ \left( {a}^{-3}-{a}^{-1} \right) {z}^{5}
 \end{array} \ee
where $a=q^{N-1}$ and $z=q-q^{-1}$.
 In agreement with
\cite{Wu} these invariants are the Kauffman polynomials \cite{K2}
of the knots.
\subsubsection{$Sp\,(2n)$ in fundamental representation}
The case of the group $Sp\,(2n)$ in fundamental representation is
very similar to $SO(2n)$ one. In appropriate basis for $N=2n$ for
universal $R$-matrix we have: \be R_{\lambda_{f}}={\hat
P}\,\sum\limits_{i\neq N+1-i}^{N} ( \,q\,e_{i\,i}\otimes
e_{i\,i}+q^{-1}e_{i\,i}\otimes e_{N+1-i\,N+1-i}\,)\,+ \atop+
\sum\limits_{i\neq j \atop i\neq N+1-j}^{N} e_{i\,i}\otimes
e_{j\,j}+\sum \limits_{i>j}^{N}\Big( (q-q^{-1})e_{i\,j}\otimes
e_{j\,i}+(q-q^{-1})\,q^{\nu_{i}-\nu_{j}} \epsilon_{i}\epsilon_{j}
\,e_{i\,j}\otimes e_{N+1-i\,N+1-j}\,\Big)\ee
 where
$$
\nu_{i}=\left\{
\begin{array}{cc}
n-i+1, & i<n+1  \\
n-i, &  i>n+1
\end{array}\right.,\ \ \ \   \epsilon_{i}=\left\{
\begin{array}{ll}
1, & i\leq n  \\
-1,&  i>n
\end{array}\right.
$$
The operator (\ref{Q}) is given by the following expression:
$$
\textsf{Q}_{\lambda_{f}}=\varrho_{\lambda_{f}}({\cal
Q})=\textrm{diag}(\,q^{\,2\,\nu_{\,2n+1-i}}\,), \ \ \ i=1,...,2n
$$
The quadratic Casimir element:
$$
\varrho_{\lambda_{f}}(\Omega_{2})=\sum\limits_{\alpha\in \Delta }
h_{\alpha^{\vee}} h_{\alpha} + \sum\limits_{\alpha\in \Phi^{+}}
e_{\alpha} f_{\alpha}+ f_{\alpha}e_{\alpha}=N-1
$$
The vev of unknot in this case:
\be<W(U_{0})>\,=\,\sum\limits_{m=1}^{2\,n} \,q^{\,2
\nu_{\,2n+1-i}} \ee and for the ambient isotopy invariant
(\ref{brep}) we get: \be \begin{array}{|c|} \hline \\
I(Sp\,(2n),\, \lambda_{f},\, K )=
\dfrac{1}{\sum\limits_{m=1}^{2\,n}\, q^{\,2\nu_{\,2n+1-i}}}
\,q^{-(N-\,1)\,w(K)}\,\tr( \textsf{Q}_{\lambda_{f}}^{\otimes n}\,
\textsf{b}_{\lambda_{f}} (K) ) \\ \\ \hline \end{array}\ee
 In this case we again obtain the Kauffman polynomial (\ref{Kauff}) for the following values
of parameters $a=-q^{2\,n+1}$ and $z=q-q^{-1}$.

\section{Conclusion}

In this paper we explicitly described the knot invariants,
constructed from the universal quantum $R$-matrix of arbitrary
simple (quantum) Lie algebra $G_q$, and demonstrated that they are
indeed invariant under all relevant Reidemeister moves. The
operators, naturally associated with this construction are,
however, the group elements of $G_q$ in the sense of \cite{MV2},
rather than the Wilson $P$-exponents in the temporal gauge. The
group elements \cite{MV2} and special point operators discussed in
this paper are made of the generators of \textit{quantum} algebra.
For this reason, one of the most important problem  is to realize
how the "primordial"  generators of a Lie algebra are transformed
into the quantum ones in the perturbation theory.  This question
and way the group elements arise as the free-field representation
of these $P$-exponents, presumably {\it a la} \cite{GMMOS},
remains to be worked out. Also relation to many other descriptions
of knot invariants, associated with other gauge choices in
Chern-Simons theory, is left beyond the scope of the present paper.\\
\indent We hope that the explicit expressions  for crossing
operators (\ref{crossings}) will help to understand better the
quantization of the CST in the temporal gauge. A possible step
toward solution of this problem is to derive the analog of the
Labastida-P\'erez (LP) formula \cite{La5}. This formula expresses
the $m$-th order of vev of Wilson loop in perturbation theory as
the trace $\tr\, T_{m}( D )$, where the element $T_{m}( D )\in
U(g)$ is defined combinatorially by the crossings of two
dimensional projection $D$ of the knot. The summation in all
orders was done in \cite{Sm} and the result was represented in the
form of contraction of the "wrong" crossing operator
(\ref{naiveR}) over all crossings. Now, we are in a reverse
situation. We know the "correct" answer for the crossing operator
(\ref{quR}) depending on $q=e^{h}$. Expanding this answer in
powers of $h$, one can find a proper analog of the LP formula.
This formula should provide some combinatorial analogs of
Kontsevich integral \cite{Konts1} for finite-type or Vassiliev
invariants \cite{Vassiliev1}. This would also help to solve a
long-standing problem of combinatorial description of Vassiliev
invariants. From the CST site, the correct analog of LP formula
would give the properly regularized perturbation theory in the
temporal gauge.
\section*{Acknowledgements}
The work was partly supported by Russian Federal Nuclear Energy
Agency, Russian Federal Agency for Science and Innovations under
the contract  02.740.5029 (A.Mor.) The work was also partly
supported by RFBR grants 09-02-00393-a(A.Mor.,A.Sm.) and
07-02-00645(A.Mor.) by joined grants RFBR-CNRS
09-01-93106(A.Mor.), 09-02-93105-CNRSL(A.Sm.),
 09-01-92440-CE(A.Mor.),
09-01-92437-CEa (A.Sm.), 09-02-91005-AFN(A.Mor.),
09-02-90493-Ukr,(A.Mor.) and by Russian president's grants for
support of scientific schools NSh-3035.2008.2 (A.Mor.) and
NSh-3036.2008.2 (A.Sm.). 
A. Smirnov is also pleased to thank the "Dynasty" foundation for support.

\end{document}